\documentclass[aps,prb,reprint,amsmath,amssymb,floatfix,nofootinbib,showkeys]{revtex4-2}

\usepackage{graphicx}
\usepackage{subfig}
\usepackage{microtype}
\usepackage{placeins}
\usepackage[percent]{overpic}
\usepackage{xcolor}
\usepackage[normalem]{ulem}

\usepackage[
  colorlinks=true,
  linkcolor=blue,
  citecolor=magenta,
  urlcolor=cyan
]{hyperref}

\begin{document}

\title{Field-Driven Coupled Magnon--Phonon--Electron Relaxation in Magnetite Using Steepest-Entropy-Ascent Quantum Thermodynamic Formalism}

\author{Deepak Dhariwal}
\affiliation{Department of Materials Science and Engineering, Virginia Tech, Blacksburg, VA 24060, USA}
\author{William T. Reynolds, Jr.}
\affiliation{Department of Materials Science and Engineering, Virginia Tech, Blacksburg, VA 24060, USA}
\author{Michael R. von Spakovsky}
\affiliation{Department of Mechanical Engineering, Virginia Tech, Blacksburg, VA 24060, USA}

\date{\today}

\begin{abstract}
A field-driven steepest-entropy-ascent quantum thermodynamic (SEAQT) formulation is developed for longitudinal nonequilibrium relaxation in magnetite (Fe$_3$O$_4$) with coupled electron, phonon, and magnon populations. Material-specific excitation spectra define the thermodynamic state space, while one relaxation parameter for each population sets its kinetic scale. A longitudinal magnetic field shifts the dressed magnon eigenenergies while the occupation basis remains fixed; irreversible redistribution conserves instantaneous energy and electron number while allowing the magnon population to vary. The formulation yields nonequilibrium subsystem temperatures, entropy production, magnetic-work identities, and a coupled small-signal susceptibility incorporating energy-conservation feedback among all three populations; the one-pole Debye response appears only as a limiting case. Numerical results under sinusoidal driving show a transition from nearly quasistatic behavior to frequency-dependent lag, finite-amplitude departure from the linear-response ellipse, and increasing higher-harmonic content. Relaxational work per cycle increases strongly with field amplitude and frequency, while the complex susceptibility is broader and shifted relative to a Debye reference. Entropy production remains positive, and the electron, phonon, and magnon temperatures show distinct excursions followed by secular heating when positive magnetic work is retained without heat rejection. The calculated work represents longitudinal magnon quasiparticle relaxation in a homogeneous single-domain model, not the total core loss of a finite ferrite specimen.
\end{abstract}


\maketitle

\section{Introduction}
\label{sec:introduction}

Soft ferrites remain indispensable magnetic-core materials for high-frequency power conversion because their comparatively high electrical resistivity suppresses the eddy-current losses that strongly constrain metallic soft magnets at elevated frequency~\cite{Stoppels1996,Thakur2020}. Their measured magnetic loss, however, is not associated with a single microscopic process. Depending on composition, microstructure, field amplitude, and frequency, experimentally inferred core loss can contain contributions from domain-wall displacement, intradomain magnetization dynamics, dielectric response, spatial current flow, and other relaxation processes~\cite{Bertotti1988,Lebourgeois1996,Tsutaoka1999,Wu2024}. Empirical descriptions such as Steinmetz-type relations are, therefore, extremely useful for engineering design, but their fitted coefficients do not provide a direct route from a material's electronic, vibrational, and magnetic excitation spectra to its non-equilibrium response~\cite{Steinmetz1892,Bertotti1988}. The objective here is narrower: to isolate the longitudinal magnon quasiparticle-relaxation contribution to energy dissipation associated with coupled electron, phonon, and magnon populations under a prescribed time-dependent magnetic field. The material dependence enters through the corresponding excitation spectra, while prescribed coarse-grained dynamic relaxation parameters set the temporal scale of the evolution.

Magnetite, Fe$_3$O$_4$, is considered as the representative material. Its inverse-spinel ferrimagnetism originates from oppositely aligned magnetic sublattices of unequal moment, whereas the mixed Fe$^{2+}$/Fe$^{3+}$ character of the octahedral sublattice gives Fe$_3$O$_4$ substantially greater electronic conductivity than many more insulating ferrites~\cite{Neel1948,Anderson1950,ZhangSatpathy1991,Degiorgi1987}. This makes Fe$_3$O$_4$ a useful test of a formulation that retains the electronic population as an active thermodynamic subsystem. The equations are otherwise material-independent since simply replacing the electronic, phonon, and magnon spectra and the associated kinetic inputs defines the corresponding model for another ferrimagnet. A subsequent comparative application of this framework examines how material-specific spectra and prescribed relaxation scales affect the dynamic response of Fe$_3$O$_4$, MnFe$_2$O$_4$, and (Mn$_{0.5}$Zn$_{0.5}$)Fe$_2$O$_4$ under consistent operating conditions~\cite{Dhariwal2026FerriteComparison}. This article treats only Fe$_3$O$_4$, for which all three excitation spectra are available from first-principle calculations~\cite{Dhariwal2026}.

The physical scope here is deliberately restricted so that a particular contribution to dissipation can be explored systematically without other convoluting factors. Namely, the approach explores how energy is dissipated when a cyclic magnetic field drives the exchange of energy through the magnon channel to the electron and phonon channels. This issue is explored as a function of the amplitude and frequency of the alternating magnetic field. To do this, it is necessary to construct a model that incorporates only those factors that contribute to the dissipation that is in view. The material is represented by one spatially homogeneous node (i.e., a single magnetic domain) with a fixed magnetic axis. The applied field and magnetization are treated as collinear scalars, and the magnetic response is represented by changes in the magnon population about one ordered branch. Domain walls, transverse precession, vector rotation, magnetocrystalline anisotropy, crystal defects, and electron transport arising from eddy currents are explicitly excluded. The resulting $M$--$H$ trajectories, therefore, represent longitudinal relaxation about a chosen ordered branch, isolated by the homogeneous single-domain model. The predictions made here are not of zero-centered major hysteresis loops or domain-switching minor loops. Likewise, the energy absorbed due to the relaxational magnetic work\footnote{This reflects the energy transfer by which the magnetization of the material returns to its equilibrium value after being disturbed due to interactions between the magnetic motion and the material's lattice structure.} density via the coupled electron--phonon--magnon channel represents only one contribution to experimentally measured core loss. If the cycle-averaged magnetic work is positive, sustained steady operation requires thermal transport to remove the accumulated excitation energy. That heat-rejection process is not included in this single-node model. This qualification is particularly important for Fe$_3$O$_4$, whose electrical conductivity is high enough that spatial charge transport cannot be presumed negligible in a general specimen without an independent geometric and electromagnetic estimate~\cite{Degiorgi1987}.

The non-equilibrium dynamics are formulated within the steepest-entropy-ascent quantum thermodynamic (SEAQT) formalism and its hypoequilibrium construction~\cite{Beretta2006,Beretta2014,Beretta2020,Li2018a,Li2018b}. A recent operator-level analysis establishes the invariant-manifold structure of hypoequilibrium states for commuting systems with fixed spectra~\cite{BerettaRayVonSpakovsky2026}. The driven magnon spectrum considered here is time dependent so that its reduced closure is derived explicitly below rather than inferred from the fixed-spectrum result. The representation follows the occupation-ladder construction developed for electron--phonon systems by Li, von Spakovsky, and Hin~\cite{Li2018b} where a continuous density of one-particle states is discretized into a finite pseudo-eigenstructure; each spectral bin is represented by an occupation ladder; and, on the hypoequilibrium manifold, the state of a ladder is described by one scalar parameter per spectral ladder. Yamada, von Spakovsky, and Reynolds show how an analogous magnon pseudo-eigenstructure can be used to describe magnetic equilibrium and non-equilibrium relaxation within the SEAQT formalism~\cite{Yamada2019}. This formulation combines these ingredients into one three-species material node and introduces a prescribed time-dependent magnetic field through the magnon Hamiltonian.

Under the single-effective-magnon-moment approximation adopted here, each represented magnon mode acquires the same field-dependent one-magnon Zeeman shift $h(t)=\mu_0\mu_mH(t)$, where $\mu_m>0$ is the effective magnetic moment removed from the ordered state by one magnon, $H(t)$ is the time-dependent magnetic field strength, and $\mu_0$ is the magnetic permeability of vacuum. The magnon one-particle energies are therefore $\varepsilon_i^m+h(t)$. Because this shift multiplies the magnon number operator, the Hamiltonians at different times are diagonal in the same occupation-number basis and commute with one another. The drive changes the eigenvalues while leaving the eigenprojectors and degeneracies fixed. This places the magnetic problem in a particularly simple commuting subset of the more general time-dependent Hamiltonian SEAQT setting~\cite{Kim2017}. The prescribed field, therefore, enters as an externally controlled work parameter of the material Hamiltonian.

The material contains three dynamical populations but only two of these contribute global dissipative constraints in the reduced description adopted here: conservation of the instantaneous dressed material energy by the irreversible redistribution of occupations and conservation of total electron number. Magnon number is not included as a conserved property. The resulting coarse-graining permits net changes in the magnon population and hence longitudinal magnetization relaxation, while microscopic angular-momentum-transfer pathways remain unresolved. This is a statement about the thermodynamic level of description, not about number conservation in every individual magnon scattering process.

Coupling among the three populations arises through the common multipliers imposed by the global constraints. The dynamic metric supplies the time scale of relaxation and may, in general, be level dependent. For the reduced calculations considered here it is represented by one prescribed effective relaxation parameter for each population, $\tau_e$, $\tau_p$, and $\tau_m$, chosen independently of the magnetic-work calculation. In particular, $\tau_m$ is interpreted as an effective \emph{longitudinal magnon-population relaxation parameter}. It is not identified automatically with a transverse Gilbert-damping or ferromagnetic-resonance time unless a separate mapping establishes that correspondence.

Several consequences follow from this construction. First, the instantaneous dressed-energy constraint yields an identity relating the change in bare excitation-energy density to the magnetic work density, $\Delta E_0=\mu_0\int H\,dM$. Second, for states initialized on the species-affine hypoequilibrium manifold, a uniform relaxation parameter within each population makes that reduced manifold invariant so that the state is described by three non-equilibrium inverse temperatures together with one additional magnon affinity generated by the time-dependent field. Thus, the three-temperature concept remains useful, but a purely three-temperature description is not generally sufficient under a changing magnetic field. Third, the entropy-production rate of the coupled material can be written as a nonnegative quadratic form in the level departures from their instantaneous constrained targets. This relation is used as a thermodynamic and numerical consistency condition rather than as an independent assumption about the magnetic work. Fourth, linearization about an equilibrium reference state at $H_b$ yields a complex susceptibility in which a direct magnon Debye-like response is modified by an energy-conservation feedback term containing the electron, phonon, and magnon spectral moments and all three relaxation parameters. A single-pole Debye response is, therefore, a controlled limiting case rather than an imposed constitutive law. Finally, without an external heat-rejection boundary, a positive value of the magnetic work density $\mu_0\int H\,dM$ over a field period raises the mean bare excitation-energy density of the node. A locally periodic magnetic oscillation must, thus, be distinguished from a strict periodic thermodynamic state.

The development presented in the following sections is limited to modeling a single crystal system in the presence of a time varying magnetic field. The focus is solely on the behavior of the electron, phonon, and magnon populations relative to each other. As a result, only a single magnetic domain is considered and no domain walls exist. The system is also assumed to be thermally isolated from its environment with no induced electrical current and, thus, no eddy losses. All of these features could be added via the SEAQT formalism but are left for future work. The development presented here, thus, begins in Section~\ref{sec:framework} with the construction of the density of states (DOS)-based state representation and the electron, phonon, and magnon pseudo-eigenstructures. The field-dressed Hamiltonian and the two-multiplier SEAQT equations of motion are then derived followed by the reduced three-temperature-plus-affinity dynamics, entropy production, and magnetic energy and work-density identities. The final part of the section develops the coupled small-signal susceptibility, its Debye and elliptical limits, and the main admissibility and consistency conditions. A spatial extension to a network of local material nodes is given separately in Appendix~\ref{app:network} so that the single-node derivation remains the central theoretical development. Section \ref{sec:numerical_implementation} then presents the numerical implementation followed by results and a discussion in Section \ref{sec:discussion}. Section \ref{sec:conclusion} provides some final conclusions.   

\section{Theoretical framework}
\label{sec:framework}

\subsection{Thermodynamic state and spectral representation}
\label{subsec:state_spectra}

The material is treated as a single spatially homogeneous region containing three populations of elementary excitations,
\begin{equation}
     k\in\{e,p,m\},
\end{equation}
for electrons, phonons, and magnons. Each population is represented by a set of one-particle energy bins $\{\varepsilon_i^k,g_i^k\}$ constructed from the continuous DOS for each of these elementary elements. The quantity $g_i^k$ denotes the number of one-particle modes (i.e., the degeneracy) represented by bin $i$. In the numerical representation, the DOS is normalized per unit of physical volume so that the sums below yield excitation-number, energy, and entropy densities. If $D_k(\varepsilon)$ denotes the DOS of species $k$, a bin bounded by $\varepsilon_{i-1/2}$ and $\varepsilon_{i+1/2}$ yields for $g_i^k>0$
\begin{align}
    g_i^k
    &=\int_{\varepsilon_{i-1/2}}^{\varepsilon_{i+1/2}}D_k(\varepsilon)\,d\varepsilon,
     \label{eq:dos_weight}\\
     \varepsilon_i^k
     &=\frac{1}{g_i^k}\int_{\varepsilon_{i-1/2}}^{\varepsilon_{i+1/2}}
     \varepsilon D_k(\varepsilon)\,d\varepsilon,
     \label{eq:dos_mean}
\end{align}
The binning need not be identical for the three species. Only the resulting level lists and their normalization must be internally consistent.

Each one-particle level is treated as an independent occupation ladder. For electrons, the occupation number is $n\in\{0,1\}$, whereas for phonons and magnons $n\in\{0,1,2,\ldots\}$. Let $p_{i,n}^k$ be the probability that one mode represented by eigenlevel $i$ of species $k$ contains $n$ excitations with
\begin{equation}
     \sum_n p_{i,n}^k=1.
     \label{eq:level_norm}
\end{equation}
At the single-level scale, the hypoequilibrium family is represented by the one-parameter distribution,
\begin{equation}
     p_{i,n}^k=\frac{\exp(-n y_i^k)}{Z_i^k},
     \label{eq:ladder_distribution}
\end{equation}
where $y_i^k$ is dimensionless and $Z_i^k = \sum_n e^{-ny_i^k}$ is the $i^{th}$ eigenlevel partition function. Using Eq.~\eqref{eq:ladder_distribution}, the expectation value for the number of particles, $\langle n \rangle_i^k = \sum_n p_{i,n}^k n$, occupying each one-particle eigenlevel $i$ is written for each elementary element as
\begin{align}
     \langle n \rangle_i^e&=\frac{1}{e^{y_i^e}+1},
     \label{eq:fermi_occ}\\
     \langle n \rangle_i^{p,m}&=\frac{1}{e^{y_i^{p,m}}-1},
     \label{eq:bose_occ}
\end{align}
where the infinite geometric series of the phonon and magnon partition function is equivalent to the closed form expression $1/(1 - e^{-y_i^{p,m}})$. Furthermore, 
\begin{equation}
     -\frac{\partial \langle n \rangle_i^k}{\partial y_i^k}=A_i^k>0,
     \label{eq:A_def}
\end{equation}
with
\begin{equation}
     A_i^e=\langle n \rangle_i^e(1-\langle n \rangle_i^e),\qquad
     A_i^{p,m}=\langle n \rangle_i^{p,m}(1+\langle n \rangle_i^{p,m}).
     \label{eq:A_explicit}
\end{equation}
The factors $A_i^k$ are occupation-fluctuation factors or equivalently the susceptibility of the mean occupation to the level affinity $y_i^k$. For electrons, $A_i^e$ vanishes in both the fully occupied and empty limits, so only thermally active states around the chemical potential contribute appreciably. For bosons the corresponding factor grows with population and emphasizes low-energy thermally occupied modes.

The excitation-number and bare-energy densities are now written as
\begin{align}
     \langle n \rangle_k&=\sum_i g_i^k \langle n \rangle_i^k,
     \label{eq:Nk}\\
     \langle e \rangle_k&=\sum_i g_i^k\varepsilon_i^k \langle n \rangle_i^k,
     \label{eq:Ek}
\end{align}
and
\begin{equation}
     \langle e \rangle_0=\langle e \rangle_e+\langle e \rangle_p+\langle e \rangle_m.
     \label{eq:E0_def}
\end{equation}
Because electron number is conserved in the model, the electronic energy zero carries the usual gauge freedom: a constant shift $\varepsilon_i^e\rightarrow\varepsilon_i^e+c$ changes $\langle e \rangle_e$ only by the constant $c\langle n \rangle_e$ and can be absorbed in the multiplier conjugate to conserved electron number $\nu\rightarrow\nu-\beta c$, which appears in the equation of motion, Eq. (\ref{eq:y_e_eom}), below. Thus, none of the occupation dynamics, energy-transfer rates, or magnetic-work-density identities depends on the arbitrary reference chosen for the electronic DOS\@. The bosonic energies do not have the same freedom because phonon and magnon numbers are not constrained globally.

The entropy density is
\begin{equation}
     \langle s \rangle =-k_B\sum_{k,i}g_i^k\sum_n p_{i,n}^k\ln p_{i,n}^k.
     \label{eq:entropy_def}
\end{equation}
with the degeneracy, $g_i^k$, interpreted as the number of independent modes represented by the bin so that Eq.~\eqref{eq:entropy_def} is simply $g_i^k$ times the entropy of the bin's representative ladder. An aggregate many-mode ladder with an explicit combinatorial multiplicity is an alternative but equivalent representation. The two descriptions should not be combined.

The three spectra entering these equations have different microscopic origins even though they enter the SEAQT state construction in the same mathematical form. The electronic DOS of Fe$_3$O$_4$ reflects the mixed-valence octahedral Fe states and the near-Fermi-level electronic structure of magnetite~\cite{ZhangSatpathy1991,Degiorgi1987,Dhariwal2026}. It is retained explicitly rather than replaced by a free-electron DOS. The phonon spectrum is obtained from first-principle lattice dynamics and contains the acoustic and optical branch structure of the spinel lattice rather than a Debye approximation~\cite{Dhariwal2026}. The magnon spectrum follows from the ferrimagnetic exchange structure and linear-spin-wave spectrum of the inequivalent $A$ and $B$ magnetic sublattices, producing multiple branches rather than the single branch of a simple ferromagnet~\cite{Yamada2019,Dhariwal2026}. Retaining these computed spectra allows the thermodynamic response weights to inherit the detailed electronic, vibrational, and magnetic structure of the material.

For the magnetic observable, each magnon is taken to reduce the ordered magnetization, $M_{sat}$, of the magnetic material by an effective magnetic moment $\mu_m>0$ such that
\begin{equation}
     M=M_{\mathrm{sat}}-\mu_m \langle n\rangle_m.
     \label{eq:M_def}
\end{equation}
where $\langle n \rangle_m$ is the expectation value of the number of magnons, which is not conserved, and $M$ the reduced magnetization. A single effective $\mu_m$ is adopted as a coarse-grained description. In a two-sublattice ferrimagnet, individual magnon branches can exhibit a different magnetic character, making a branch- or level-resolved $\mu_i^m$ a natural refinement~\cite{Nambu2020}. Nonetheless, one effective moment is retained here to isolate the thermodynamic coupling problem from the additional branch-polarization bookkeeping. In numerical applications, the effective $\mu_m$ must be chosen consistent with the magnetic structure used to construct the magnon spectrum. In addition, $M_{\mathrm{sat}}$ is only required when absolute, rather than reference-subtracted, magnetization is reported.

The harmonic spin-wave description also defines a physical validity range for the magnon pseudo-eigenstructure. It treats excitations about an ordered ferrimagnetic reference state and does not by itself include strong magnon--magnon renormalization, collapse of the ordered moments, or critical fluctuations close to the Curie temperature. Outside this regime, the limitation lies in the supplied harmonic magnon spectrum: the SEAQT evolution does not restore physics that is absent from the underlying pseudo-eigenstructure. To do so would require using an anharmonic magnom spectrum, which accounts for the additional physics.

\subsection{Field-driven SEAQT dynamics and three-species coupling}
\label{subsec:driven_equations}

The prescribed local longitudinal magnetic field strength acting on the modeled node is
\begin{equation}
     H(t)=H_b+H_0\cos\omega t,
     \label{eq:H_drive}
\end{equation}
where $\omega$ is the angular frequency of the magnetic field, $H_b$ the strength of a bias field, and $H_0$ the alternating-current amplitude. Within the single-effective-moment approximation introduced above, the one-magnon Zeeman shift energy is defined as
\begin{equation}
     h(t)=\mu_0\mu_m H(t).
     \label{eq:h_def}
\end{equation}
where $\mu_0$ is the magnetic permeability. Multiplying Eq.~\eqref{eq:M_def} by $-\mu_0H(t)$, the magnetic energy of the material in the presence of the field becomes
\begin{equation}
     -\mu_0HM=-\mu_0HM_{\mathrm{sat}}+h\langle n\rangle_m.
\end{equation}
The first term on the right is independent of the magnon occupations, whereas the second is occupation dependent. Since $\langle n\rangle_m = \sum_i g_i^m \langle n\rangle_i^m$, the latter term can be written as $h\langle n\rangle_m = \sum_i g_i^m\, h\, \langle n\rangle_i^m$. Combining this with the bare magnon energy $\langle e\rangle_m = \sum_i g_i^m\, \varepsilon_i^m\, \langle n\rangle_i^m$ yields the dressed magnon energy $\langle e\rangle_m^{\rm dressed} = \sum_i g_i^m\, (\varepsilon_i^m + h)\, \langle n\rangle_i^m$, which indicates that the field shifts every one-magnon eigenenergy by the same amount so that 
\begin{equation}
     \epsilon_i^m(t)=\varepsilon_i^m+h(t),
     \label{eq:dressed_magnon}
\end{equation}
The eigenenergies $\epsilon_i^e = \varepsilon_i^e$ and $\epsilon_i^p = \varepsilon_i^p$ for the electrons and phonons are unaffected.

The time-dependent excitation Hamiltonian operator, $\hat{\mathsf{H}}_m^{\mathrm{exc}}(t)$, for the occupation-dependent magnons can now be written as 
\begin{equation}
\begin{aligned}
     \hat{\mathsf{H}}_m^{\mathrm{exc}}(t) &= \sum_i g_i^m\,\epsilon_i^m(t)\hat n_i^m,\\
     &= \sum_i g_i^m\,\varepsilon_i^m \hat n_i^m + h(t) \sum_i g_i^m\,\hat n_i^m \\
     &= \hat{\mathsf{H}}_{m, 0} + h(t) \hat{\mathsf{N}}_m
\end{aligned}
 \label{eq:Hm}
\end{equation}
while the magnetic operator is expressed as 
\begin{equation}
     \hat{\mathsf{M}} = M_{\mathrm{sat}}\hat{\mathsf{I}}-\mu_m\hat{\mathsf{N}}_m
    \end{equation}
so that
\begin{align}
     -\mu_0 H \hat{\mathsf{M}} &= -\mu_0 H M_{\mathrm{sat}}\hat{\mathsf{I}} + \mu_0 H \mu_m\hat{\mathsf{N}}_m \nonumber \\
     -\mu_0 H \hat{\mathsf{M}} &= -\mu_0 H M_{\mathrm{sat}}\hat{\mathsf{I}} + h\hat{\mathsf{N}}_m
     \label{eq:M_operator}
\end{align}
 Note that $\mu_0 H \hat M$ is not the energy stored in the magnetic field itself. It is instead the (Zeeman) interaction energy transferred to the material from the magnetic field. The full field-dependent material Hamiltonian operator can now be written as
\begin{align}
     \hat{\mathsf H}(t)&=\hat{\mathsf H}_0-\mu_0H(t)\hat{\mathsf{M}} \nonumber \\
     &=\hat{\mathsf{H}}_0+h(t)\hat{\mathsf{N}}_m-\mu_0H(t)M_{\mathrm{sat}}\hat{\mathsf{I}}.
     \label{eq:full_Hamiltonian}
\end{align}
where $\hat{\mathsf{H}}_0$ is the zero-field effective quasiparticle Hamiltonian operator constructed from the electron, phonon, and magnon pseudo-eigenstructures obtained from the corresponding first-principles and spin-wave calculations. The last term in Eq.~\eqref{eq:full_Hamiltonian} is proportional to the identity operator and, therefore, affects the field-dependent energy offset and external work but not the occupation redistribution. In the harmonic quasiparticle representation adopted here, $[\hat{\mathsf{H}}_0,\hat{\mathsf{N}}_m]=0$. Because the field-dependent part contains only $\hat{\mathsf{ N}}_m$ and the identity,
\begin{equation}
     [\hat{\mathsf H}(t),\hat{\mathsf H}(t')]=0,\qquad
     [\hat{\mathsf H}(t),\dot{\hat{\mathsf H}}(t)]=0.
     \label{eq:commuting_H}
\end{equation}
The field, thus, changes eigenvalues while leaving the instantaneous eigenprojectors fixed. This commuting structure is a special case of the general time-dependent Hamiltonian SEAQT treatment of Kim and von Spakovsky~\cite{Kim2017}. In that treatment, the dissipative contribution to the time-dependent energy balance contains a term expressible through the matrix elements of the commutator $[\hat{\mathcal H},\dot{\hat{\mathcal H}}]$. For present case, the commutator-dependent correction appearing in the general energy balance vanishes (see Eq.~\eqref{eq:commuting_H}). The dissipative redistribution is, therefore, constrained to conserve the instantaneous dressed energy density, while explicit variation of the Hamiltonian supplies energy via a work interaction at the rate $\mathrm{Tr}(\rho\dot{\hat{\mathsf H}})=-\mu_0M\dot H$.

The irreversible SEAQT redistribution is constrained by normalization of each occupation ladder, conservation of total electron number, and conservation of instantaneous dressed material energy by the dissipative part of the motion. Magnon number is not included among the global constraints. Introducing
\begin{equation}
     \Delta_i^k=y_i^k-\beta \epsilon_i^k-\nu\delta_{ke},
     \label{eq:Delta_def}
\end{equation}
where $\delta_{ke}$ is the Kronecker delta,  $\beta(t)$ the common multiplier conjugate to the dressed energy and $\nu(t)$ the multiplier conjugate to the conserved electron number. With one effective relaxation parameter per species, the equations of motion for each species can be written as
\begin{align}
     \dot y_i^e&=-\frac{1}{\tau_e}
     (y_i^e-\beta\varepsilon_i^e-\nu),
     \label{eq:y_e_eom}\\
     \dot y_i^p&=-\frac{1}{\tau_p}
     (y_i^p-\beta\varepsilon_i^p),
     \label{eq:y_p_eom}\\
     \dot y_i^m&=-\frac{1}{\tau_m}
     [y_i^m-\beta \epsilon_i^m(t)].
     \label{eq:y_m_eom}
\end{align}
Using Eq.~\eqref{eq:A_def}, the equation for the time rate of change of $\langle n\rangle_i^k $ for each species is expressed as 
\begin{equation}
     \langle \dot n \rangle_i^k=\frac{A_i^k}{\tau_k}\Delta_i^k.
     \label{eq:n_eom}
\end{equation}
A level-resolved dynamic metric replaces $\tau_k$ by $\tau_i^k$. Species-uniform times are needed only for the reduced affine closure derived in Sec.~\ref{subsec:temperature_entropy_energy}, not for the instantaneous multiplier solution itself.

A positive response weight, which will be used below, is now defined as
\begin{equation}
     r_i^k\equiv\frac{g_i^kA_i^k}{\tau_k}.
     \label{eq:response_weight}
\end{equation}
It combines the number of represented modes, their instantaneous occupation susceptibility, and their kinetic rate and indicates that each eigenlevel contributes strongly to the coupled constraints only when it is simultaneously numerous, thermodynamically active, and assigned a comparatively rapid kinetic rate. Furthermore, electron-number conservation requires that
\begin{equation}
     \sum_i r_i^e\Delta_i^e=0,
     \label{eq:constraint_N}
\end{equation}
and the conservation of the instantaneous dressed energy requires that
\begin{equation}
     \sum_{k,i}r_i^k \epsilon_i^k\Delta_i^k=0.
     \label{eq:constraint_E}
\end{equation}
Now, using the positive response weight and the following definitions for the response-weighted Gram-matrix elements, $Q_{\alpha \gamma}$, of the energy and electron-number constraints, and the corresponding projections, $R_{\alpha}$, of the instantaneous eigenlevel coordinates $y_i^k$ where the subscripts $\alpha, \gamma \in \{ E, N\}$:
\begin{align}
     Q_{EE}&=\sum_{k,i}r_i^k(\epsilon_i^k)^2,\;
     Q_{EN}=\sum_i r_i^e\varepsilon_i^e, \;
     Q_{NN}&=\sum_i r_i^e, \\
     R_E&=\sum_{k,i}r_i^k \epsilon_i^k y_i^k,\;
     R_N=\sum_i r_i^e y_i^e,
     \label{eq:Q_definitions}
\end{align}
the system of equations which determine the two global multipliers, $\beta$ and $\nu$, appearing in Eqs. (\ref{eq:y_e_eom}) to (\ref{eq:n_eom}) are written as
\begin{equation}
     \boxed{
     \begin{pmatrix}
     Q_{EE}&Q_{EN}\\Q_{EN}&Q_{NN}
     \end{pmatrix}
     \begin{pmatrix}\beta\\\nu\end{pmatrix}
     =\begin{pmatrix}R_E\\R_N\end{pmatrix}}.
     \label{eq:two_by_two_solve}
\end{equation}
This $2\times2$ system is the central coupling mechanism of the single-node model. The field enters through the dressed magnon energies and the magnon occupations, all three species contribute to the energy constraint, and only the electron block contributes to the number constraint. The determinant of this matrix is given by
\begin{equation}
     \mathcal D=Q_{EE}Q_{NN}-Q_{EN}^2.
     \label{eq:determinant}
\end{equation}
Defining the response-weighted electron mean energy $\bar\varepsilon_e=Q_{EN}/Q_{NN}$ and, of course, assuming $Q_{NN}>0$, the determinant can be written in the form
\begin{equation}
     \begin{aligned}
     \mathcal D=Q_{NN}\Bigg[&\sum_i r_i^e(\varepsilon_i^e-\bar\varepsilon_e)^2\\
     &+\sum_i r_i^p(\varepsilon_i^p)^2+\sum_i r_i^m[\epsilon_i^m (t)]^2\Bigg].
     \end{aligned}
     \label{eq:determinant_positive}
\end{equation}
Hence $\mathcal D\ge0$ and is strictly positive when the electron-number constraint has nonzero response weight and at least one independent energy direction is thermodynamically active. If $Q_{NN}=0$, no active electronic degree of freedom remains and the electron-number constraint is removed, leaving the corresponding energy-only problem.

The relaxation parameters in Eq.~\eqref{eq:response_weight} are dynamic inputs, not quantities determined uniquely by the three DOS\@. In the more general SEAQT construction the relaxation metric can be level dependent, and its connection to dynamic and scattering information is discussed explicitly in the electron--phonon formulation of Li \textit{et al.}~\cite{Li2018b}. The present $\tau_e$, $\tau_p$, and $\tau_m$ should, therefore, be read as coarse-grained species-level representations of unresolved microscopic scattering rates. They set the physical time scale on which the SEAQT state-space trajectory is traversed and consequently the frequency windows in which the different populations can follow the drive. They are prescribed independently of the loss curves predicted here rather than fitted to those curves. Experimental relaxation measurements can motivate their orders of magnitude, but they do not define unique, process-independent species-level values.

\subsection{Species temperatures, entropy production, and magnetic energy balance}
\label{subsec:temperature_entropy_energy}

The eigenlevel equations of Sec.~\ref{subsec:driven_equations} provide the general dynamical description. A lower-dimensional species-affine closure is available when the initial state lies on the corresponding hypoequilibrium manifold, describing a material prepared in mutual equilibrium or in a species-wise hypoequilibrium state before the field protocol begins, and when all levels of a given species share the same relaxation parameter. For fixed spectra, the invariant-manifold character of hypoequilibrium states and the relaxation law of their intensive parameters have been established within earlier SEAQT constructions~\cite{Li2018a,Li2018b} and, more recently, in operator form~\cite{BerettaRayVonSpakovsky2026}.

Because the magnon spectrum here depends explicitly on time, the driven closure is obtained directly below by coefficient matching rather than by leaving the fixed-spectrum result unchanged. For the preparation considered here the appropriate affine forms are
\begin{align}
     y_i^e&=\beta_e\varepsilon_i^e+\nu_e,
     \label{eq:affine_e}\\
     y_i^p&=\beta_p\varepsilon_i^p,
     \label{eq:affine_p}\\
     y_i^m&=\beta_m \epsilon_i^m(t)+\eta_m.
     \label{eq:affine_m}
\end{align}
The quantities $\beta_e$, $\beta_p$, and $\beta_m$ are non-equilibrium inverse-temperature coordinates defined in terms of each population temperature, $T_k$, where  
\begin{equation}
     T_k=\frac{1}{k_B\beta_k}.
     \label{eq:species_temperature}
\end{equation}
The electron coordinate $\nu_e$ is required by the conservation of electron number, which at equilibrium corresponds to the usual chemical-potential, $\mu_e$, combination $\nu_e=-\beta_e\mu_e$ under the sign convention of Eq.~\eqref{eq:fermi_occ}. The additional magnon coordinate $\eta_m$ is a level-independent non-equilibrium affinity measured relative to the instantaneous field-dressed magnon spectrum. It is not a multiplier enforcing conservation of total magnon number but instead is required to preserve the affine form under the time-dependent shift $\epsilon_i^m(t)$\footnote{A more general phonon affine form could contain a level-independent intercept $\eta_p$, obeying $\dot\eta_p=-\eta_p/\tau_p$. For the equilibrium-prepared phonon population considered here, $\eta_p(0)=0$ and remains so and is, thus, omitted without approximation for that initial manifold.}.

Substitution of Eqs.~\eqref{eq:affine_e} and \eqref{eq:affine_p} into the electron and phonon eigenlevel equations of motion, Eqs. (\ref{eq:y_e_eom}) and (\ref{eq:y_p_eom}), and matching coefficients of the eigenlevel energies yields
\begin{align}
     \dot\beta_e&=-\frac{\beta_e-\beta}{\tau_e},
     &\dot\nu_e&=-\frac{\nu_e-\nu}{\tau_e},
     \label{eq:betae_nue_eom}\\
     \dot\beta_p&=-\frac{\beta_p-\beta}{\tau_p}.
     \label{eq:betap_eom}
\end{align}
For magnons, differentiating Eq.~\eqref{eq:affine_m} results in
\begin{equation}
     \dot y_i^m=\dot\beta_m \epsilon_i^m+\beta_m\dot h+\dot\eta_m.
\end{equation}
Comparing this last expression with Eq.~\eqref{eq:y_m_eom} provides the equations for $\dot \beta_m$ and $\dot \eta_m$, i.e.,
\begin{align}
     \dot\beta_m&=-\frac{\beta_m-\beta}{\tau_m},
     \label{eq:betam_eom}\\
     \dot\eta_m&=-\frac{\eta_m}{\tau_m}-\beta_m\dot h.
     \label{eq:eta_eom}
\end{align}
Eqs.~\eqref{eq:betae_nue_eom}--\eqref{eq:eta_eom} establish the reduced invariant manifold for species-uniform relaxation parameters. The driven affine manifold therefore contains three inverse-temperature coordinates and one additional magnon affinity, generated by the $-\beta_m\dot h$ term. In the static-field limit, $\dot h=0$ and $\eta_m$ relaxes exponentially to zero, recovering a pure three-temperature form on the dressed spectrum.

The common multiplier $\beta$ can be written explicitly in terms of the variables of the reduced invariant manifold. Defining the response-weighted moments
\begin{align}
     W_j^e&=\sum_i r_i^e(\varepsilon_i^e)^j,
     &W_j^p&=\sum_i r_i^p(\varepsilon_i^p)^j,
     \label{eq:W_ep}\\
     \widetilde W_j^m&=\sum_i r_i^m[\epsilon_i^m(t)]^j,
     \label{eq:W_m_dressed}
\end{align}
for $j=0,1,2$, where the tilde indicates that the magnon response-weighted moment is based on the dressed as opposed to bare magnon energy. The electron fixed-number energy variance can then be written as
\begin{equation}
     V_e=W_2^e-\frac{(W_1^e)^2}{W_0^e}\ge0,
     \label{eq:Ve}
\end{equation}
while the electron-number constraint is expressed as
\begin{equation}
     \nu=\nu_e+(\beta_e-\beta)\frac{W_1^e}{W_0^e}.
     \label{eq:nu_reduced}
\end{equation}
Using this result in the dressed-energy constraint leaves an electron contribution $V_e(\beta_e-\beta)$, a phonon contribution $W_2^p(\beta_p-\beta)$, and a magnon contribution
\begin{equation}
     \widetilde W_2^m(\beta_m-\beta)+\eta_m\widetilde W_1^m.
\end{equation}
Hence
\begin{equation}
     \boxed{
     \beta=
     \frac{V_e\beta_e+W_2^p\beta_p+\widetilde W_2^m\beta_m
     +\eta_m\widetilde W_1^m}
     {V_e+W_2^p+\widetilde W_2^m}}.
     \label{eq:beta_three_temp_affinity}
\end{equation}
When $\eta_m=0$ this reduces to the fluctuation- and rate-weighted three-temperature average
\begin{equation}
     \boxed{
     \beta=
     \frac{V_e\beta_e+W_2^p\beta_p+\widetilde W_2^m\beta_m}
     {V_e+W_2^p+\widetilde W_2^m}}.
     \label{eq:beta_three_temp}
\end{equation}
The affinity term is essential under a time-dependent field since it shifts the instantaneous composite multiplier away from the value determined by the three species temperatures alone.

The reduced equations also recover the familiar two-temperature structure as a controlled limit. If the magnon block is removed from the coupling and $\eta_m=0$, Eq.~\eqref{eq:beta_three_temp} reduces to
\begin{equation}
     \beta=\frac{V_e\beta_e+W_2^p\beta_p}{V_e+W_2^p}.
     \label{eq:beta_two_temp}
\end{equation}
Substitution into Eqs.~\eqref{eq:betae_nue_eom} and \eqref{eq:betap_eom} results in 
\begin{align}
     \dot\beta_e
     &=-\frac{1}{\tau_e}
     \frac{W_2^p}{V_e+W_2^p}(\beta_e-\beta_p),
     \label{eq:two_temp_e}\\
     \dot\beta_p
     &=-\frac{1}{\tau_p}
     \frac{V_e}{V_e+W_2^p}(\beta_p-\beta_e).
     \label{eq:two_temp_p}
\end{align}
The coupling coefficients are controlled by the relative fluctuation-weighted energy measures of the two subsystems. In the limit that one subsystem carries a much larger active energy weight than the other, its inverse temperature changes comparatively little, while the smaller subsystem relaxes predominantly toward it. This is the same two-temperature structure recovered for electron--phonon coupling in the SEAQT formulation of Li \textit{et al.}~\cite{Li2018b}, now appearing as a limiting case of the present three-species construction.

The entropy-production rate follows from the same set of eigenlevel equations of motion. For a one-parameter occupation ladder, differentiation of Eq.~\eqref{eq:entropy_def} gives the hypoequilibrium relation~\cite{Li2018b}
\begin{equation}
     \frac{1}{k_B}\langle \dot s\rangle_i^k=y_i^k \langle \dot n\rangle_i^k.
     \label{eq:level_entropy_rate}
\end{equation}
Using Eq.~\eqref{eq:n_eom} and summing over all of the eigenlevels and the three species, the entropy production rate for the material, which in this case is equivalent to the time rate of change of the entropy, $\langle \dot s \rangle$, of the system, is written as
\begin{equation}
     \frac{\langle\dot s\rangle}{k_B}=\sum_{k,i}r_i^k y_i^k\Delta_i^k.
     \label{eq:Sdot_intermediate}
\end{equation}
Since
\begin{equation}
     y_i^k=\Delta_i^k+\beta \epsilon_i^k+\nu\delta_{ke},
\end{equation}
the dressed-energy and electron-number constraints remove the final two terms exactly leaving
\begin{equation}
     \boxed{
     \frac{\langle\dot s\rangle}{k_B}=\sum_{k,i}r_i^k(\Delta_i^k)^2\ge0}.
     \label{eq:entropy_production}
\end{equation}
No linear-response approximation is used in Eq.~\eqref{eq:entropy_production}. The explicit field dependence changes the Hamiltonian energy but, in the fixed-projector representation of Eq.~\eqref{eq:commuting_H}, does not itself change the von Neumann entropy. This is as it should be since the magnetic field interaction with the material is that of a work interaction, which does not involve the transfer of entropy. The entropy changes ony because of the irreversible occupation redistribution represented by Eq.~\eqref{eq:n_eom}. On the reduced manifold and using Eq.~\eqref{eq:nu_reduced}, Eq. (\ref{eq:Delta_def}) can be expressed for each species as
\begin{align}
     \Delta_i^e&=(\beta_e-\beta)
     \left(\varepsilon_i^e-\frac{W_1^e}{W_0^e}\right),
     \label{eq:Delta_e_reduced}\\
     \Delta_i^p&=(\beta_p-\beta)\varepsilon_i^p,
     \label{eq:Delta_p_reduced}\\
     \Delta_i^m&=(\beta_m-\beta)\epsilon_i^m+\eta_m.
     \label{eq:Delta_m_reduced}
\end{align}
Substituting into Eq. (\ref{eq:entropy_production}) then yields
\begin{equation}
     \boxed{
     \begin{aligned}
     \frac{\langle \dot s \rangle}{k_B}
     &=V_e(\beta_e-\beta)^2+W_2^p(\beta_p-\beta)^2\\
     &\quad+\sum_i r_i^m\left[(\beta_m-\beta)\epsilon_i^m+\eta_m\right]^2
     \end{aligned}}.
     \label{eq:entropy_production_reduced}
\end{equation}
This expression separates the electron fixed-number variance, the phonon thermal departure, and the combined thermal--affinity departure of the driven magnon block. Eq.~\eqref{eq:entropy_production_reduced} therefore provides both the thermodynamic measure of irreversible redistribution and an independent consistency identity for the dynamics. It is not used as a substitute for the independent magnetic-work balance.

The corresponding energy-density exchange rates among the populations follow directly from the reduced departures. The electron and phonon bare-energy-density rates are
\begin{align}
     \langle \dot e\rangle_e&=V_e(\beta_e-\beta),
     \label{eq:Ee_rate_reduced}\\
     \langle\dot e\rangle_p&=W_2^p(\beta_p-\beta),
     \label{eq:Ep_rate_reduced}
\end{align}
whereas the magnon contribution to the occupation part of the dressed energy-density rate is
\begin{equation}
     \langle \dot e\rangle_m = \sum_i g_i^m \epsilon_i^m\langle\dot n\rangle_i^m
     =\widetilde W_2^m(\beta_m-\beta)+\eta_m\widetilde W_1^m.
     \label{eq:Em_dressed_rate_reduced}
\end{equation}
Eq.~\eqref{eq:beta_three_temp_affinity} is precisely the condition for which the sum of Eqs.~\eqref{eq:Ee_rate_reduced}, \eqref{eq:Ep_rate_reduced}, and \eqref{eq:Em_dressed_rate_reduced} vanish. Thus the same algebra that fixes the common temperature target also enforces instantaneous energy exchange among the three populations consistent with energy conservation.

For the magnetic work density, return to the bare excitation-energy density $\langle e\rangle_0$ in Eq.~\eqref{eq:E0_def}. The dressed-energy constraint is equivalent to
\begin{equation}
     \sum_i g_i^e\varepsilon_i^e\dot n_i^e
     +\sum_i g_i^p\varepsilon_i^p\dot n_i^p
     +\sum_i g_i^m[\varepsilon_i^m+h(t)]\dot n_i^m=0.
     \label{eq:dressed_energy_rate_constraint}
\end{equation}
Hence
\begin{equation}
     \langle \dot e\rangle_0+h\langle\dot n\rangle_m=0.
     \label{eq:E0_Nm}
\end{equation}
Using $\dot M=-\mu_m\langle\dot n\rangle_m$ gives
\begin{equation}
     \boxed{\langle\dot e\rangle_0=\mu_0H\dot M},
     \label{eq:E0_work_rate}
\end{equation}
and, over any interval,
\begin{equation}
     \boxed{
     \langle e\rangle_0(t_2)-\langle e\rangle_0(t_1)=\mu_0\int_{t_1}^{t_2}H\,dM}.
     \label{eq:E0_work_integral}
\end{equation}
This identity does not require periodicity.

The full field-dependent material energy density, which includes the Zeeman term, is written a
\begin{equation}
     \langle e\rangle_{fd}=\langle e\rangle_0-\mu_0HM.
     \label{eq:U_def}
\end{equation}
Taking the total derivative of this last expression and using Eq.~\eqref{eq:E0_work_rate} results in
\begin{equation}
     \boxed{\langle \dot e\rangle_{fd} =-\mu_0M\dot H}.
     \label{eq:U_work_rate}
\end{equation}
The two equivalent work-density forms are therefore related by
\begin{equation}
     -\mu_0\int_{t_1}^{t_2}M\,dH
     =\mu_0\int_{t_1}^{t_2}H\,dM
     -\mu_0[HM]_{t_1}^{t_2}.
     \label{eq:integration_by_parts}
\end{equation}
For a closed $H$--$M$ cycle, and more generally whenever the endpoint product $HM$ is unchanged, the boundary term vanishes and the two work-density integrals coincide. This distinction matters because the present single-node model contains no heat sink. For a drifting trajectory, $\mu_0\int H\,dM$ equals the change in the bare excitation-energy density, whereas the work density associated with the full Hamiltonian is $-\mu_0\int M\,dH$. The two differ by the endpoint term in Eq.~\eqref{eq:integration_by_parts}.

Entropy production and magnetic work density thus describe different aspects of the same driven evolution. Eq.~\eqref{eq:entropy_production} quantifies irreversible redistribution in state space, while Eq.~\eqref{eq:E0_work_integral} quantifies energy density transferred to the bare excitations by the prescribed field. Because the node is not held at an externally imposed constant temperature, there is no general exact identity of the form $W=T\Delta \langle s\rangle$ over a finite driven cycle. Any such proportionality would require an additional near-isothermal approximation or an explicit thermal boundary condition.

\subsection{Linear response, Debye reduction, and nonlinear departure}
\label{subsec:linear_response}

The nonlinear eigenlevel equations govern the finite-amplitude response. Their small-signal limit is nevertheless important because it yields a closed analytical susceptibility, which provides an analytical benchmark for the nonlinear calculation, and because it separates linear multi-timescale coupling from genuinely nonlinear loop distortion. A periodic first-order response is compatible with the absence of a finite-loss thermal limit cycle because net energy absorption begins at $O(H_0^2)$. The $O(H_0)$ oscillation can, therefore, be periodic about a reference state, while the mean energy drifts only on the slower second-order scale.

Consider an admissible equilibrium reference state at temperature $T_0$ and bias field $H_b$, with
\begin{equation}
     \beta_0=(k_BT_0)^{-1},\qquad
     \bar h=\mu_0\mu_mH_b,\qquad
     \bar \epsilon_i^m=\varepsilon_i^m+\bar h.
     \label{eq:reference_state}
\end{equation}
At equilibrium
\begin{align}
     y_i^{e,0}&=\beta_0\varepsilon_i^e+\nu_0,\\
     y_i^{p,0}&=\beta_0\varepsilon_i^p,\\
     y_i^{m,0}&=\beta_0\bar \epsilon_i^m,
\end{align}
and $\eta_m=0$. Let a perturbation vary as $e^{i\omega t}$, with
\begin{equation}
     h(t)=\bar h+\delta \breve h(t),\qquad
     \delta \breve h=\mu_0\mu_m\delta \breve H.
\end{equation}
where $\delta\breve h $  and $\delta\breve H$ are complex. To first order, variations of $A_i^k=A_i^{k,0}$ (see Eq. (\ref{eq:A_explicit})) do not enter the constraint coefficients because the reference departures $\Delta_i^{k,0}$ vanish at equilibrium.

Linearizing the eigenlevel equations yields
\begin{align}
     (1+i\omega\tau_e)\delta \breve y_i^e
     &=\varepsilon_i^e\delta\breve\beta+\delta\breve\nu,
     \label{eq:lin_y_e}\\
     (1+i\omega\tau_p) \delta \breve y_i^p
     &=\varepsilon_i^p\delta\breve\beta,
     \label{eq:lin_y_p}\\
     (1+i\omega\tau_m)\delta \breve y_i^m
     &=\bar \epsilon_i^m\delta\breve \beta
     +\beta_0\delta \breve h.
     \label{eq:lin_y_m}
\end{align}
Defining
\begin{align}
     B_{NN}^e&=\sum_i g_i^eA_i^{e,0},\;
     B_{EN}^e=\sum_i g_i^eA_i^{e,0}\varepsilon_i^e,\nonumber \\
     B_{EE}^e&=\sum_i g_i^eA_i^{e,0}(\varepsilon_i^e)^2,\\
     C_e&=B_{EE}^e-\frac{(B_{EN}^e)^2}{B_{NN}^e},
     \label{eq:electron_linear_moments}\\
     B_{EE}^p&=\sum_i g_i^pA_i^{p,0}(\varepsilon_i^p)^2,
     \label{eq:phonon_linear_moment}\\
     B_{NN}^m&=\sum_i g_i^mA_i^{m,0},\;
     B_{EN}^m=\sum_i g_i^mA_i^{m,0}\bar \epsilon_i^m,
     \nonumber\\
     B_{EE}^m&=\sum_i g_i^mA_i^{m,0}(\bar \epsilon_i^m)^2,
     \label{eq:magnon_linear_moments}
\end{align}
the electron-number conservation is expressed as
\begin{equation}
     \delta\breve\nu
     =-\frac{B_{EN}^e}{B_{NN}^e}\delta\breve\beta.
     \label{eq:delta_nu}
\end{equation}
The linearized dressed-energy constraint can then be written in terms of
\begin{equation}
     \boxed{
     G(\omega)=
     \frac{C_e}{1+i\omega\tau_e}
     +\frac{B_{EE}^p}{1+i\omega\tau_p}
     +\frac{B_{EE}^m}{1+i\omega\tau_m}}.
     \label{eq:Gomega}
\end{equation}
so that
\begin{equation}
     \boxed{
     \delta\breve \beta
     =-\frac{\beta_0\delta \breve h\,B_{EN}^m}
     {(1+i\omega\tau_m)G(\omega)}}.
     \label{eq:delta_beta}
\end{equation}
The derivation of this last expression is given in Appendix \ref{app:susceptibility_derivation}. The common energy multiplier, therefore, oscillates even though the external drive enters explicitly only through the magnon energy. This is the linear expression of the feedback by which the driven magnon population communicates with the phonon and electron blocks through the common energy constraint.

Since $\delta \breve n_i^m=-A_i^{m,0}\delta \breve y_i^m$ and $\delta \breve M=-\mu_m\delta \breve n_m$,
\begin{equation}
     \delta \breve M
     =\mu_m\sum_i g_i^mA_i^{m,0}\delta \breve y_i^m.
     \label{eq:delta_M_from_y}
\end{equation}
Writing $\delta \breve M=\chi(\omega)\delta \breve H$ where $\chi(\omega)$ is the susceptibility yields the following equation for $\chi(\omega)$:
\begin{equation}
     \boxed{
     \chi(\omega)=\mu_0\mu_m^2\beta_0
     \left[
     \frac{B_{NN}^m}{1+i\omega\tau_m}
     -\frac{(B_{EN}^m)^2}
     {(1+i\omega\tau_m)^2G(\omega)}
     \right]}.
     \label{eq:coupled_chi}
\end{equation}
where again the derivation of this last expression is found in Appendix \ref{app:susceptibility_derivation}. The first term in the square brackets is the direct longitudinal magnon-population response. The second is the feedback required by the dressed-energy conservation. Via $G(\omega)$, the feedback contains the thermodynamically active parts of all three spectra and all three relaxation parameters. Multi-timescale or non-Debye frequency dependence can, therefore, appear even while the response remains strictly linear in field amplitude.

When the feedback term is small compared with the direct magnon contribution,
\begin{equation}
     \left|\frac{(B_{EN}^m)^2}
     {(1+i\omega\tau_m)G(\omega)}\right|\ll B_{NN}^m,
     \label{eq:debye_condition}
\end{equation}
Eq.~\eqref{eq:coupled_chi} reduces to
\begin{equation}
     \chi(\omega)\rightarrow\frac{\chi_{\mathrm D,0}}{1+i\omega\tau_m},\qquad
     \chi_{\mathrm D,0}=\mu_0\mu_m^2\beta_0B_{NN}^m.
     \label{eq:debye_limit}
\end{equation}
Thus, a Debye law is recovered as a limiting form of the coupled susceptibility.

With
\begin{equation}
     \chi(\omega)=\chi'(\omega)-i\chi''(\omega),\qquad \chi''\ge0,
     \label{eq:chi_convention}
\end{equation}
a sinusoidal perturbation $\delta \breve H=H_0\cos\omega t$ gives
\begin{equation}
     \delta \breve M=H_0[\chi'\cos\omega t+\chi''\sin\omega t].
     \label{eq:linear_M_time}
\end{equation}
Eliminating time yields
\begin{equation}
     |\chi|^2(\delta \breve H)^2-2\chi'\delta \breve H\,\delta \breve M+(\delta \breve M)^2
     =(\chi''H_0)^2,
     \label{eq:ellipse}
\end{equation}
which is an ellipse whenever $\chi''\ne0$. This conclusion is independent of whether $\chi(\omega)$ follows the single-pole Debye form. A non-Debye frequency dependence can, therefore, occur entirely within a linear response, whereas departure from an elliptical single-tone $M$--$H$ trajectory signals finite-amplitude nonlinearity or loss of periodic closure.

Now, the second-order relaxational work density per cycle follows from Eq.~\eqref{eq:E0_work_integral},
\begin{equation}
     \boxed{
     W_{\mathrm{rel}}^{(2)}=\mu_0\oint H\,dM
     =\pi\mu_0H_0^2\chi''(\omega)}.
     \label{eq:linear_loss}
\end{equation}
where the superscript (2) indicates second order. At finite amplitude, the full Fermi/Bose maps, the state-dependent factors $A_i^k$, and the multiplier solution make the governing equations nonlinear even though $H(t)$ is sinusoidal. The magnetization can then contain higher harmonics such that
\begin{equation}
     M(t)=M_0+\sum_{n\ge1}
     [M_n'\cos(n\omega t)+M_n''\sin(n\omega t)].
     \label{eq:harmonic_expansion}
\end{equation}
These higher, i.e., Fourier, harmonics, distort the $M$--$H$ trajectory away from an ellipse, but for a strictly periodic response under the single-tone field of Eq.~\eqref{eq:H_drive} they do not contribute independently to the net loop work density. Orthogonality of the Fourier harmonics gives the exact closed-cycle identity
\begin{equation}
     \boxed{
     W_{\mathrm{cyc}}=\mu_0\oint H\,dM
     =\pi\mu_0H_0M_1''
     =\pi\mu_0H_0^2\chi_1''(H_0,\omega)},
     \label{eq:nonlinear_first_harmonic_work}
\end{equation}
where $\chi_1''(H_0,\omega)\equiv M_1''/H_0$ is the amplitude-dependent quadrature coefficient of the fundamental, i.e., of the lowest frequency component of the periodic waveform. Eq.~\eqref{eq:nonlinear_first_harmonic_work} reduces to the linear-response result of Eq.~\eqref{eq:linear_loss} as $H_0\rightarrow0$, for which $\chi_1''\rightarrow\chi''(\omega)$. Higher harmonics, therefore, quantify waveform distortion but do not contribute independently to the net closed-cycle work density under a pure sinusoidal drive. Suitable measures of nonlinear response include the amplitude dependence of the fundamental, harmonic ratios such as $|M_3|/|M_1|$, and geometric departure of the trajectory from its best first-harmonic ellipse. Equality between the direct closed-loop integral and the fundamental quadrature expression provides an independent periodic-closure check.

\subsection{Limiting behavior, admissibility, and consistency relations}
\label{subsec:limits_validation}

Several limiting cases provide physical interpretation and stringent checks on the formulation. In the absence of a time-dependent field ($\dot h=0$), Eq.~\eqref{eq:eta_eom} drives $\eta_m\rightarrow0$. The reduced state then approaches a three-temperature form on the static dressed spectrum. If the three populations remain mutually coupled and no additional conserved quantity prevents equilibration, Eqs.~\eqref{eq:betae_nue_eom}--\eqref{eq:betam_eom} drive $\beta_e$, $\beta_p$, and $\beta_m$ toward a common inverse temperature while Eq.~\eqref{eq:nu_reduced} enforces the conserved electron number. A change of the prescribed field performs work through the explicit Hamiltonian dependence, Eq.~\eqref{eq:U_work_rate}. Subsequent relaxation redistributes that energy among the coupled populations. In a thermally isolated node, the resulting equilibrium temperature can, therefore, change with the field, providing the thermodynamic setting for a magnetocaloric response.

In the quasistatic limit\footnote{The ``quasistatic limit'' results from the series of final stable equilibrium states to which the system evolves via the SEAQT equation of motion for a single initial state when the frequency is zero and the equation of motion is repeatedy executed for different magnetic field strengths, $H$.}, the drive period is long compared with the relevant relaxation parameters and the state remains close to the instantaneous constrained-equilibrium manifold. The irreversible departures $\Delta_i^k$ become small, Eq.~\eqref{eq:entropy_production} approaches zero, and the forward and reverse magnetic trajectories retrace as the irreversible departures vanish. Accordingly,
\begin{equation}
 \lim_{\omega\rightarrow0}\mu_0\oint H\,dM=0
 \label{eq:quasistatic_loss_zero}
\end{equation}
for a closed quasistatic field cycle that remains on the same ordered branch. The instantaneous temperature need not be constant during that cycle. In a thermally isolated material, it can vary reversibly with field. What vanishes is the irreversible loop area, not necessarily the magnetocaloric response.

At the opposite extreme, if $\omega\tau_k\gg1$ for the populations carrying the response, the occupation distributions cannot follow the rapidly changing target during one period. Eq.~\eqref{eq:coupled_chi} then suppresses the amplitude of the magnetic response, and the closed-cycle relaxational work tends toward zero even though the corresponding power need not have the same asymptotic dependence. The frequency-dependent phase lag and relaxational work are, therefore, concentrated in intermediate regimes where one or more of $\omega\tau_e$, $\omega\tau_p$, and $\omega\tau_m$ are not asymptotically small or large. Because the three channels are coupled through $G(\omega)$, the locations and number of spectral features need not coincide with the simple conditions $\omega\tau_k=1$.

A further distinction is necessary because the primary single-node model contains no heat-rejection boundary. Defining the bare-excitation increment accumulated over one field period by
\begin{equation}
     \Delta \langle e \rangle_{0,\mathrm{cyc}}
     \equiv\mu_0\int_t^{t+T}H\,dM.
 \label{eq:positive_cycle_work}
\end{equation}
shows the change in bare excitation-energy density over the period even when the state does not close. If $\Delta \langle e \rangle_{0,\mathrm{cyc}}>0$, the beginning thermodynamic state cannot repeat exactly after every cycle. A driven calculation can nevertheless enter a regime in which the within-cycle oscillation changes only slowly from one period to the next. Such a trajectory is better described as a locally periodic response superposed on a slow thermal drift than as an exact limit cycle.

The accumulated bare excitation-energy density can also be mapped to an equivalent equilibrium temperature at fixed bias field and electron number. Let $\langle e \rangle_{0,\mathrm{eq}}(T,H_b,\langle n\rangle_e)$ be the equilibrium \emph{bare excitation-energy density} at the bias field and conserved electron number. The corresponding bare-energy temperature derivative, $C_0$, is then written as
\begin{equation}
     C_0(T,H_b,\langle n\rangle_e)
     =\left(\frac{\partial \langle e\rangle_{0,\mathrm{eq}}}{\partial T}\right)_{H_b,\langle n\rangle_e}.
     \label{eq:C0}
\end{equation}
$C_0$ is distinguished from the heat capacity associated with the full field-dependent Hamiltonian, whose energy also contains the Zeeman term. For a sufficiently small increment,
\begin{equation}
     \Delta T_{\mathrm{cyc}}\simeq\frac{\Delta \langle e\rangle_{0,\mathrm{cyc}}}{C_0},
     \label{eq:deltaT_approx}
\end{equation}
whereas a finite increment should be mapped by solving
\begin{equation}
     \langle e\rangle_{0,\mathrm{eq}}(T+\Delta T,H_b,\langle n\rangle_e)
     =\langle e\rangle_{0,\mathrm{eq}}(T,H_b,\langle n\rangle_e)+\Delta \langle e\rangle_{0,\mathrm{cyc}}.
     \label{eq:deltaT_exact}
\end{equation}
This construction is an energy-based equilibrium mapping. It is not an additional definition of the instantaneous non-equilibrium species temperatures in Eq.~\eqref{eq:species_temperature}. Predicting a laboratory temperature when the material is thermally interacting with its surroundings requires an additional thermal boundary model, sample geometry, and heat-transfer information not contained in the three quasiparticle spectra and three local relaxation parameters. This could be modeled within the SEAQT formalism and the hypoequilibrium description by adding an additional subsystem to act as a thermal reservoir but was not included since it was beyond the intended scope of the present work.

As to the bosonic sector, it imposes an independent admissibility condition. The actual magnon ladder requires that
\begin{equation}
     y_i^m(t)>0
     \qquad\text{for every represented magnon bin},
     \label{eq:y_positive}
\end{equation}
so that the Bose partition sum remains finite. A sufficient condition for the instantaneous canonical target itself to remain admissible at positive temperature is
\begin{equation}
     \boxed{
     \epsilon_i^m(t)=\varepsilon_i^m+\mu_0\mu_mH(t)>0
     \quad\forall i,t}.
     \label{eq:boson_target_admissibility}
\end{equation}
For the sinusoidal protocol of Eq.~\eqref{eq:H_drive}, a convenient conservative condition is that
\begin{equation}
     \varepsilon_{\min}^m+\mu_0\mu_m(H_b-H_0)>0.
     \label{eq:bias_condition}
\end{equation}
Here $\varepsilon_{\min}^m$ denotes the lowest sampled positive zero-field magnon energy represented in the discretized pseudo-eigenstructure.  The underlying exchange-only spin-wave spectrum need not possess a strict anisotropy-induced gap; consequently, the lowest sampled positive energy is not interpreted here as evidence of such a physical gap.  Because the magnon density of states becomes extremely small in this low-energy region, however, the same quantity is used in Sec.~\ref{sec:numerical_implementation} as an effective low-energy spectral-onset scale,
\begin{equation}
    \varepsilon_{\rm gap}
    \equiv
    \varepsilon_{\min}^m,
    \label{eq:effective_gap}
\end{equation}
for nondimensionalizing the applied-field amplitude. The nonlinear calculations parameterize the AC-field amplitude through the dimensionless quantity
\begin{equation}
    \xi
    \equiv
    \frac{\mu_0\mu_m H_0}{\varepsilon_{\rm gap}},
    \label{eq:xi_definition}
\end{equation}
where $\varepsilon_{\rm gap}$ is the lowest sampled positive magnon energy in the discretized magnon spectrum used to construct the SEAQT pseudo-eigenstructure.  Because the magnon density of states is extremely small in the low-energy region approaching this value, $\varepsilon_{\rm gap}$ is used here as an effective low-energy spectral-onset scale.  It is not interpreted as a strict anisotropy-induced gap of the underlying exchange-only spin-wave Hamiltonian.  Equation~\eqref{eq:xi_definition} may equivalently be written as
\begin{equation}
    H_0
    =
    \frac{\xi\,\varepsilon_{\rm gap}}
    {\mu_0\mu_m},
\end{equation}
so that increasing $\xi$ corresponds directly to increasing the physical amplitude of the imposed longitudinal field.

For the zero-bias nonlinear calculations, $H_b=0$.  Using Eq.~\eqref{eq:xi_definition}, Eq.~\eqref{eq:bias_condition} then reduces to the conservative condition
\begin{equation}
    \boxed{\xi<1}.
    \label{eq:xi_admissibility}
\end{equation}
The maximum drive amplitude used in the present nonlinear calculations, $\xi=0.85$, therefore remains below the point at which the lowest
field-dressed magnon pseudo-level would reach zero energy.  

Admissibility of the actual trajectory additionally requires Eq.~\eqref{eq:y_positive} at all times. Eq.~\eqref{eq:bias_condition} constrains only the instantaneous canonical target and guarantees that it does not cross the bosonic divergence. The lowest nonzero bin of a discretized magnon DOS is not, by itself, evidence of a physical anisotropy gap because its value can depend on wave-vector sampling and binning. Until a physical gap is established independently, the conservative regime is a longitudinal protocol that satisfies Eq.~\eqref{eq:bias_condition} and remains on the chosen ordered branch.

Finally, the single-node model requires the three spectral density of states and three dynamic relaxation parameters, i.e., 
\begin{equation}
     \{D_e(E),D_p(E),D_m(E);\tau_e,\tau_p,\tau_m\}
     \label{eq:min_inputs}
\end{equation}
together with a consistent spectral-density convention; the conserved electron number or equivalent equilibrium Fermi-level initialization, $\mu_m$; and $M_{\mathrm{sat}}$ if absolute magnetization is required. The field $H(t)$ in Eq.~\eqref{eq:H_drive} is the local longitudinal field at the modeled node. Conversion from an externally applied laboratory field to this local field may require demagnetizing and, in a spatially varying problem, electromagnetic corrections determined by specimen geometry. In addition, several identities provide independent numerical consistency tests of an implementation. Electron number and the instantaneous dressed-energy constraint should be conserved to integration tolerance. The entropy rate evaluated from Eq.~\eqref{eq:level_entropy_rate} should agree with the quadratic form in Eq.~\eqref{eq:entropy_production}, while finite changes in $\langle e\rangle_0$ and $\langle e\rangle_{fd}$ should agree with $\mu_0\int H\,dM$ and $-\mu_0\int M\,dH$, respectively, including the endpoint term in Eq.~\eqref{eq:integration_by_parts} when the trajectory is not closed. In the small-signal limit, the first harmonic provides a direct check on Eq.~\eqref{eq:coupled_chi}. For a closed periodic trajectory under a single-tone drive, the direct loop integral must also agree with Eq.~\eqref{eq:nonlinear_first_harmonic_work}. These thermodynamic identities are consistency conditions, not substitutes for ordinary numerical convergence tests of the time integration and spectral discretization.

\section{Numerical implementation}
\label{sec:numerical_implementation}

The equations in Sec.~\ref{sec:framework} were implemented in Python using the DOS-based pseudo-eigenstructure described in Sec.~\ref{subsec:state_spectra}. The electronic, phonon and magnon spectra are the first-principles Fe$_3$O$_4$ inputs associated with Ref.~\cite{Dhariwal2026}. The spectra are treated only as fixed material inputs to the SEAQT state construction. Their first-principles generation is not part of the present numerical study.

The calculations in this manuscript are initialized from an equilibrium reference state at the stated temperature before the sinusoidal field protocol is applied. Electron number is conserved, while phonon and magnon numbers are allowed to evolve according to the global constraint structure derived in Sec.~\ref{subsec:driven_equations}. For the species-uniform kinetic model, one effective relaxation parameter is prescribed for each excitation population. The reference set used throughout the calculations is
\begin{equation}
    (\tau_e,\tau_p,\tau_m)=(0.05,\,3,\,500)~\mathrm{ps}.
    \label{eq:numerical_tau_set}
\end{equation}
These quantities are species-level coarse-grained dynamic inputs to the SEAQT metric. They are neither generated by the electronic, phonon, or magnon DOS nor fitted to the calculated magnetic-work or susceptibility curves. They should also not be interpreted as unique directly measured material constants. Rather, the values in Eq.~\eqref{eq:numerical_tau_set} are chosen baseline estimates representing well-separated electronic, lattice, and longitudinal magnetic relaxation scales. Time-resolved measurements on Fe$_3$O$_4$ demonstrate that electronic, lattice-associated, and magnetic recovery processes can indeed occupy widely separated time windows, extending from sub-100-fs dynamics to hundreds of picoseconds~\cite{Lu2022,Hsia2009}. Those measurements provide physical support for the separated kinetic hierarchy adopted here, but their experimentally fitted decay constants are not mapped one-to-one onto the SEAQT parameters $\tau_e$, $\tau_p$, and $\tau_m$. In particular, $\tau_m$ denotes an effective longitudinal magnon-population relaxation paramter in the present scalar model and is not identified with a transverse Gilbert-damping or ferromagnetic-resonance relaxation time.

The nonlinear calculations use the dimensionless drive amplitude $\xi$ defined in Eq.~\eqref{eq:xi_definition}.  Larger $\xi$ corresponds to a larger imposed AC-field amplitude $H_0$. The magnetic observable is evaluated from the magnon population through Eq.~\eqref{eq:M_def}. For a trajectory that closes over a field period, the relaxational work density is evaluated as
\begin{equation}
 W_{\mathrm{cyc}}=\mu_0\oint H\,dM,
 \label{eq:numerical_work}
\end{equation}
with the finite-interval identity in Eq.~\eqref{eq:integration_by_parts} retained when the thermodynamic state drifts between the beginning and end of the sampled period. Finite-amplitude distortion is characterized both geometrically, by departure of the $M$--$H$ trajectory from its first-harmonic ellipse, and spectrally through the ratio $|M_3|/|M_1|$. The small-signal response is represented by the complex susceptibility $\chi=\chi'-i\chi''$ and compared with the one-pole Debye limit of Eq.~\eqref{eq:debye_limit}. The nonequilibrium population temperatures are obtained from Eq.~\eqref{eq:species_temperature}. Under a changing field, they are supplemented by the independent magnon affinity $\eta_m$.

The entropy calculation uses the SEAQT production rate of Eq.~\eqref{eq:entropy_production}. Fig.~\ref{fig:entropy_production} reports both a cycle-integrated entropy-production measure and a rate measure in the plotting normalization shown on the axes. These two plotted ordinates are treated as distinct outputs of the calculation. Their absolute vertical scales are not converted from one to the other in the analysis below.

The numerical conditions represented in the figures are as follow. The 300 K frequency-dependent $M$--$H$ family contains a quasistatic reference together with 500~kHz, 1, 10, 100, 300, and 500~MHz, 1~GHz, and 5~GHz. The finite-amplitude $M$--$H$ and harmonic analyses at 300~K use 10, 100, 300, and 500~MHz, 1~GHz, and 5~GHz with $\xi=0.001$, 0.01, 0.05, 0.1, 0.3, 0.5, 0.7, and 0.85. The work and entropy plots use the same six frequencies with $\xi=0.01$, 0.1, 0.5, and 0.85. The temperature-dependent susceptibility is evaluated at 250, 300, 325, 350, 375, and 400~K. The subsystem-temperature plots use a frequency sweep at $\xi=0.85$, an amplitude sweep at 300~MHz, and a cycle-number evolution at 300~MHz and $\xi=0.85$.

\section{Discussion}
\label{sec:discussion}

The numerical results probe four linked consequences of the field-driven construction: the emergence of a frequency-dependent longitudinal phase lag, the breakdown of the first-harmonic linear description at finite amplitude, the associated magnetic work and entropy production, and the redistribution of the absorbed energy among the electron, phonon, and magnon populations. These quantities should not be read as independent diagnostics. Within the present formulation they are different projections of the same constrained state-space SEA evolution: the field acts directly on the magnon energies, the common energy multiplier couples the three populations, and the resulting occupation redistribution determines both the magnetic response and the irreversible thermodynamic evolution. The interpretation remains within the scope established in Sec.~\ref{sec:introduction}. In particular, the $M$--$H$ trajectories below are not domain-switching hysteresis loops, and the work represented by their longitudinal lag is not the total measured core loss of a finite ferrite specimen.

\subsection{Frequency-dependent longitudinal response and the linear benchmark}
\label{subsec:discussion_frequency}

\begin{figure*}[!t]
    \centering
    \includegraphics[width=0.82\textwidth]{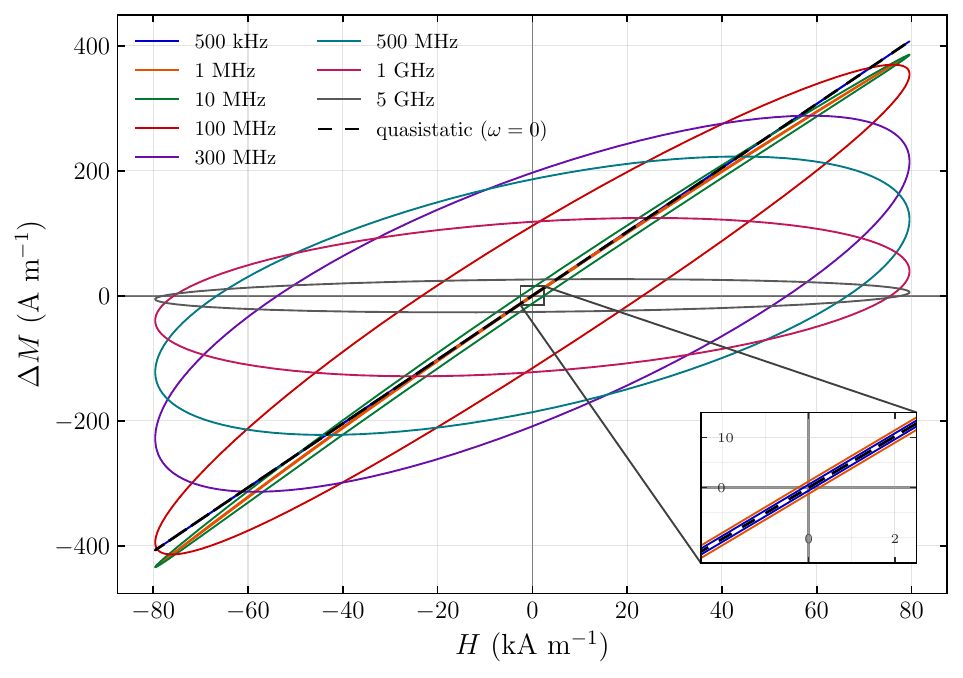}
    \caption{Frequency-dependent longitudinal $M$--$H$ response at 300~K plotted as $M-M_{\mathrm{ref}}$. The dashed curve is the quasistatic reference. The field excursion is approximately $\pm80~\mathrm{kA\,m^{-1}}$. The calculated trajectories represent longitudinal quasiparticle-population lag within one ordered branch and should not be interpreted as domain-switching major hysteresis loops.}
    \label{fig:mh_frequency}
\end{figure*}

Fig.\ref{fig:mh_frequency} shows the calculated longitudinal $M$--$H$ response at 300~K over more than four decades of drive frequency. The 500-kHz, 1-MHz, and 10-MHz trajectories remain close to the quasistatic response on the scale of the full plot. Substantial opening develops through the 100--500-MHz range, together with a clear phase lag between field and magnetization. At still higher frequency, the magnetization excursion is reduced, consistent with the occupation variables becoming progressively less able to follow the changing constrained target within one period. This sequence agrees with the two asymptotic limits derived in Sec.~\ref{subsec:limits_validation}: the irreversible loop area vanishes as $\omega\rightarrow0$, whereas sufficiently rapid driving suppresses the response amplitude. The broad intermediate regime is, therefore, the physically relevant window for relaxational loss in the present model, but it cannot be identified with one relaxation pole merely from the opening of the loop because all three populations contribute to the constrained dynamics.

\begin{figure*}[!t]
    \centering
    \subfloat[]{\includegraphics[width=0.47\textwidth]{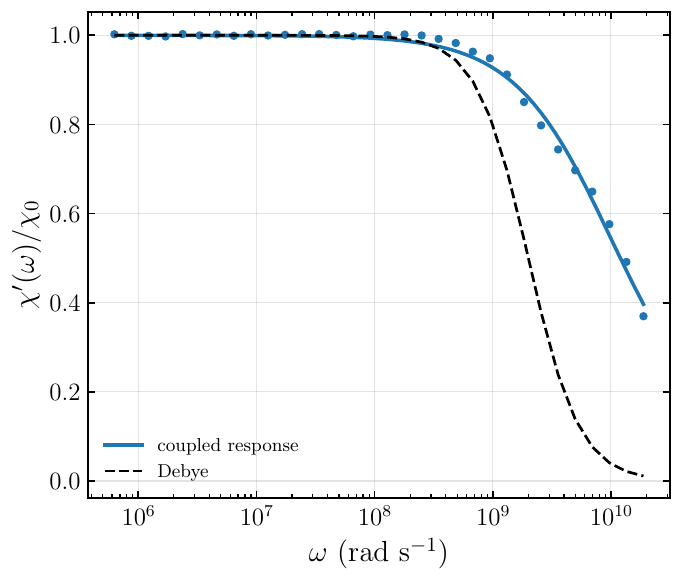}}
    \hfill
    \subfloat[]{\includegraphics[width=0.47\textwidth]{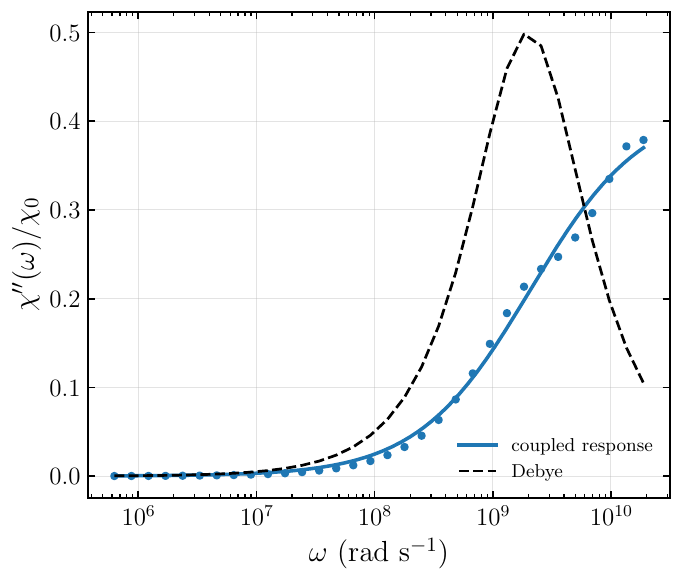}}
    \caption{Normalized small-signal susceptibility. Panels (a) and (b) show the in-phase and out-of-phase components, respectively, for the coupled SEAQT response and the one-pole Debye reference. The ordinate axis is normalized with respect to $\chi_0$. The comparison emphasizes the displacement and broadening of the coupled dispersion rather than an absolute static-amplitude comparison.}
    \label{fig:normalized_chi}
\end{figure*}

The loop geometry can be connected directly to the linear-response variables. For a purely sinusoidal response, Eq.~\eqref{eq:ellipse} shows that $\chi'$ controls the in-phase projection of the trajectory, while $\chi''$ controls its quadrature opening. Eq.~\eqref{eq:linear_loss} then makes the enclosed area proportional to $H_0^2\chi''$. Thus, the opening of the time-domain trajectories in Fig.~\ref{fig:mh_frequency} and the growth of the quadrature response in Fig.~\ref{fig:normalized_chi} are not separate phenomena. They are the time-domain and frequency-domain representations of the same lag in the linear regime. The eventual contraction of the $M$ excursion at the highest plotted frequencies is equally important: a large phase delay does not by itself imply arbitrarily large work if the response amplitude is simultaneously being suppressed.

The small-signal susceptibility in Fig.~\ref{fig:normalized_chi} makes the distinction from a one-pole Debye law more explicit. The ordinate is reported using the $\chi_0$ normalization supplied with the figure. The comparison, therefore, emphasizes the frequency-dependent line shape rather than identifying the normalization with the analytical Debye-limit amplitude $\chi_{\mathrm D,0}$ in Eq.~\eqref{eq:debye_limit}. The coupled $\chi'$ remains close to its low-frequency value to a higher angular frequency than the Debye reference and then decreases more gradually. Correspondingly, the coupled $\chi''$ is broader and its maximum is displaced toward higher angular frequency. Eq.~\eqref{eq:coupled_chi} identifies the origin of this broadening within the model. The direct magnon term is modified by the feedback denominator $G(\omega)$, which contains the thermodynamically active electron, phonon, and magnon energy moments and their respective relaxation parameters. The response can, therefore, be distributed over several coupled dynamical scales even though the perturbation remains strictly linear in field amplitude.


This distinction also clarifies how the present calculation should be compared with magnetic-response measurements. Frequency-, field-, and temperature-dependent complex susceptibility has been measured in magnetite in very different experimental regimes, from low-frequency single-crystal measurements to microwave permeability measurements of Fe$_3$O$_4$ powders and particle-based systems~\cite{Ozdemir2009,Williams2016}. Those experiments establish that magnetite can display substantial dispersive and dissipative magnetic responses, but they do not constitute a direct quantitative validation of Fig.~\ref{fig:normalized_chi} since domain-wall motion, magnetocrystalline anisotropy, finite-particle rotation, ferromagnetic resonance, demagnetizing fields, and spatial electromagnetic effects may contribute in the experimental specimens and are deliberately absent here. The useful comparison is, therefore, qualitative and mechanistic. The present result isolates the part of the dispersion that follows from longitudinal quasiparticle redistribution subject to the common SEAQT constraints.

The separation between non-Debye dispersion and nonlinearity is consequently essential. Eq.~\eqref{eq:coupled_chi} can depart strongly from a Debye line shape, while remaining exactly first order in the applied field. Evidence for finite-amplitude nonlinearity must instead come from amplitude dependence, non-elliptical single-tone trajectories, or harmonic content beyond the fundamental.

\subsection{Finite-amplitude departure from the linear-response ellipse}
\label{subsec:discussion_nonlinear}

\begin{figure*}[!t]
    \centering
    \includegraphics[width=\textwidth]{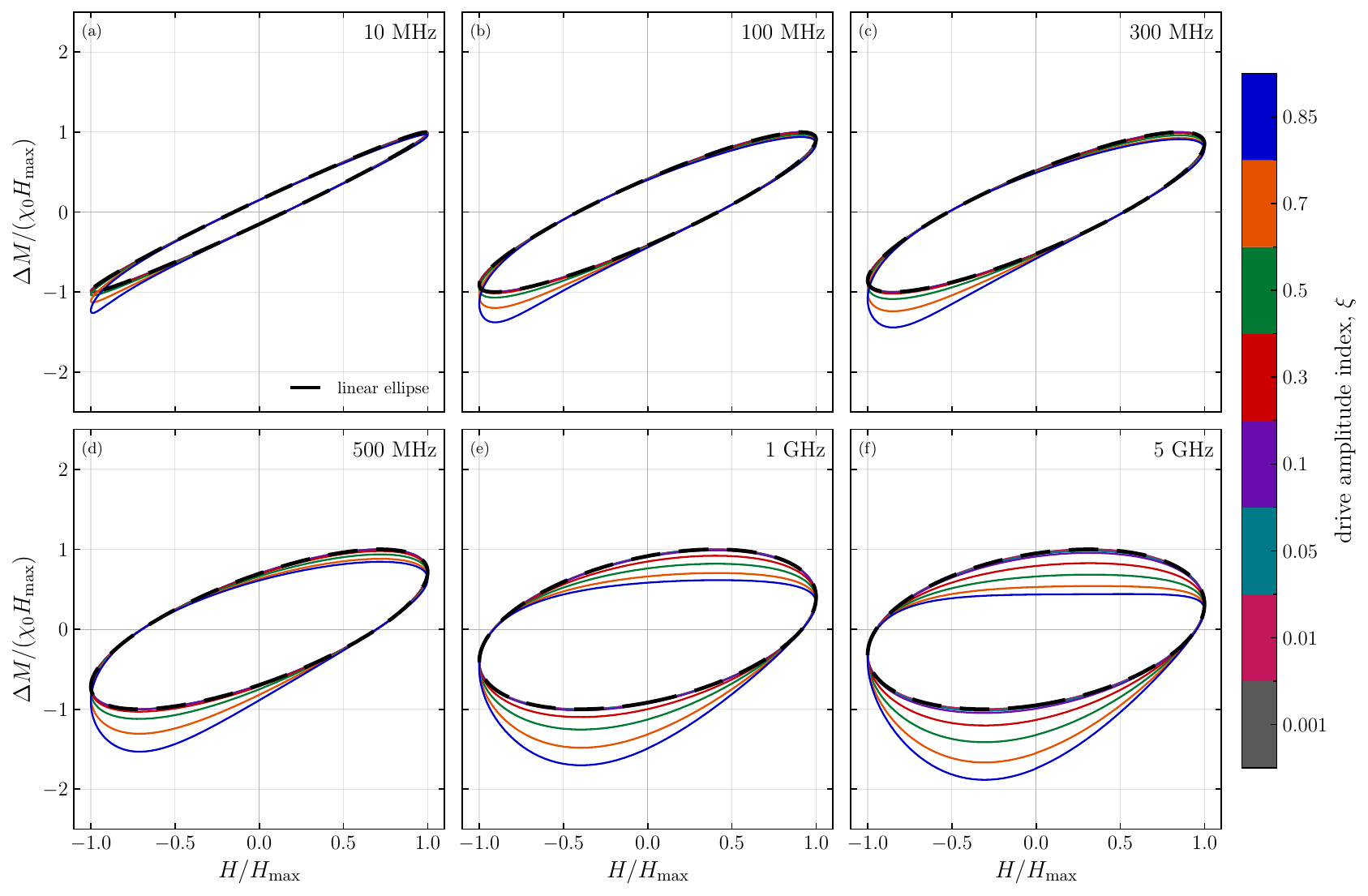}
    \caption{Finite-amplitude longitudinal $M$--$H$ trajectories at 300~K. Each panel (a) - (f) compares the normalized nonlinear trajectory with its first-harmonic linear ellipse over the indicated frequency. The dimensionless drive-amplitude index $\xi$ increases from 0.001 to 0.85. Departure from the ellipse provides a geometric measure of nonlinear response.}
    \label{fig:nonlinear_loops}
\end{figure*}

Fig.\ref{fig:nonlinear_loops} compares the finite-amplitude trajectories with their first-harmonic ellipses. At the smallest values of $\xi$, the calculated curves follow the linear ellipse closely. Because both axes are normalized by the imposed field amplitude and the fundamental magnetization scale, this agreement is more informative than an unnormalized visual overlap since the dominant trivial change in scale has been divided out. The progressively larger noncollapse that remains as $\xi$ increases, therefore, reflects a change in waveform shape rather than merely a larger magnetization excursion. The deformation is especially visible in the higher-frequency panels, where finite-amplitude state excursions occur, while the populations are already unable to relax quasistatically during each period.

The origin of this nonlinear shape change is contained in the full level equations rather than in an added hysteresis prescription. The Fermi and Bose occupations are nonlinear functions of the level affinities, while the fluctuation factors $A_i^k$ change with the instantaneous state and the global multipliers are recomputed from the state-dependent constraint matrix. A sinusoidal field can, therefore, generate a magnetization that is not sinusoidal. The additional magnon affinity $\eta_m$ is also important conceptually since under a changing field the magnon state cannot, in general, be represented by a temperature coordinate alone. The finite-amplitude response consequently samples both thermal and affinity departures from the instantaneous dressed target.

The third-harmonic ratio in Fig.~\ref{fig:harmonic_ratio} supplies an independent spectral measure of the same departure. Across all six frequencies, $|M_3|/|M_1|$ rises with increasing $\xi$, eventually by several orders of magnitude. For a zero-centered, single-tone drive, the third harmonic is a particularly useful low-order indicator because an approximately antisymmetric longitudinal magnetization response naturally emphasizes odd-order distortion. More generally, higher-order alternating current susceptibility is a standard diagnostic of nonlinear magnetic response in ordered magnetic materials~\cite{Canepa2007}. Here, the diagnostic is applied specifically to the longitudinal population dynamics generated by the SEAQT equations rather than to domain or precessional nonlinearities.

The frequency dependence of the third-harmonic growth also shows that drive amplitude and relaxation cannot be treated as independent corrections. At a given $\xi$, the state reached during a half-cycle depends on the redistribution that can occur on that time scale, while at fixed frequency, the state dependence of the occupation factors becomes progressively more important as the excursion grows. Amplitude and frequency, therefore, enter the nonlinear response jointly. At the smallest displayed amplitudes, the extremely small third-harmonic values should not be overinterpreted point by point. The physically secure result is the systematic growth of harmonic content away from the linear limit.

Higher harmonics alter the trajectory geometry but do not constitute independent additive work channels for a strictly periodic response to a single-tone field. Eq.~\eqref{eq:nonlinear_first_harmonic_work} shows that the closed-cycle work is determined by the quadrature component of the fundamental. This suggests a useful finite-amplitude bridge between the harmonic and energetic descriptions since an effective fundamental quadrature coefficient $\chi_{1,\mathrm{eff}}''(H_0,\omega)=M_1''/H_0$ can be defined for which $W_{\mathrm{cyc}}=\pi\mu_0H_0^2\chi_{1,\mathrm{eff}}''$ whenever the trajectory closes. The higher harmonics then quantify how strongly the waveform has departed from the linear constitutive form, whereas the fundamental quadrature retains the direct connection to the net closed-cycle work.

\subsection{Relaxational work, entropy production, and the thermodynamic meaning of loss}
\label{subsec:discussion_dissipation}

\begin{figure*}[!t]
    \centering
    \begin{minipage}[t]{0.48\textwidth}
    \centering
    \includegraphics[width=\linewidth]{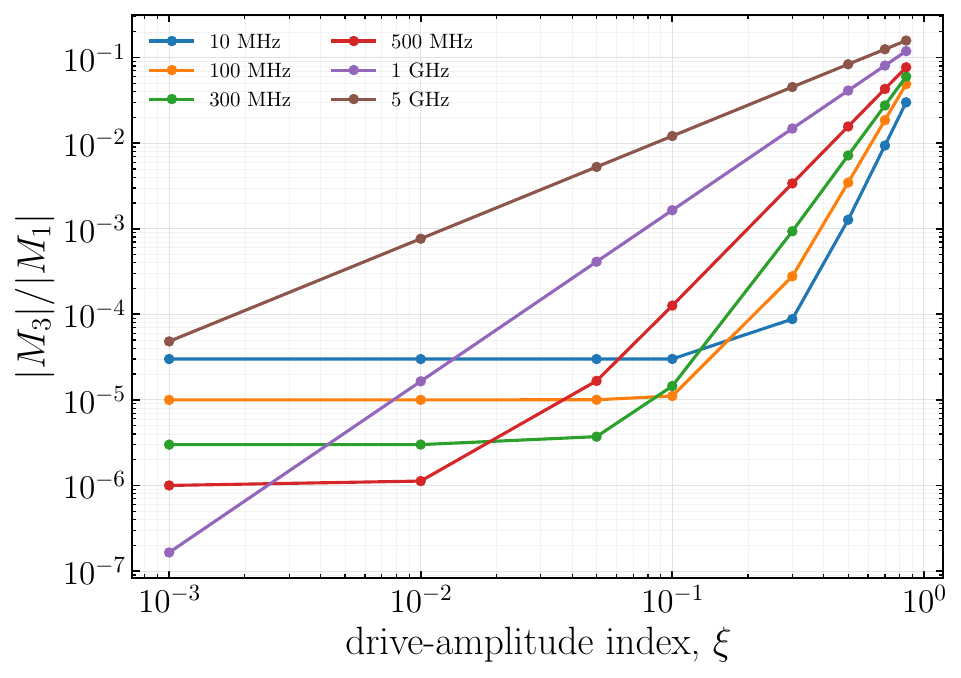}
    \caption{Third-to-first harmonic ratio $|M_3|/|M_1|$ as a function of the dimensionless drive-amplitude index $\xi$ for six frequencies. The increase with amplitude is the spectral counterpart of the loop-shape deformation in Fig.~\ref{fig:nonlinear_loops}.}
    \label{fig:harmonic_ratio}
    \end{minipage}
    \hfill
    \begin{minipage}[t]{0.48\textwidth}
    \centering
    \includegraphics[width=\linewidth]{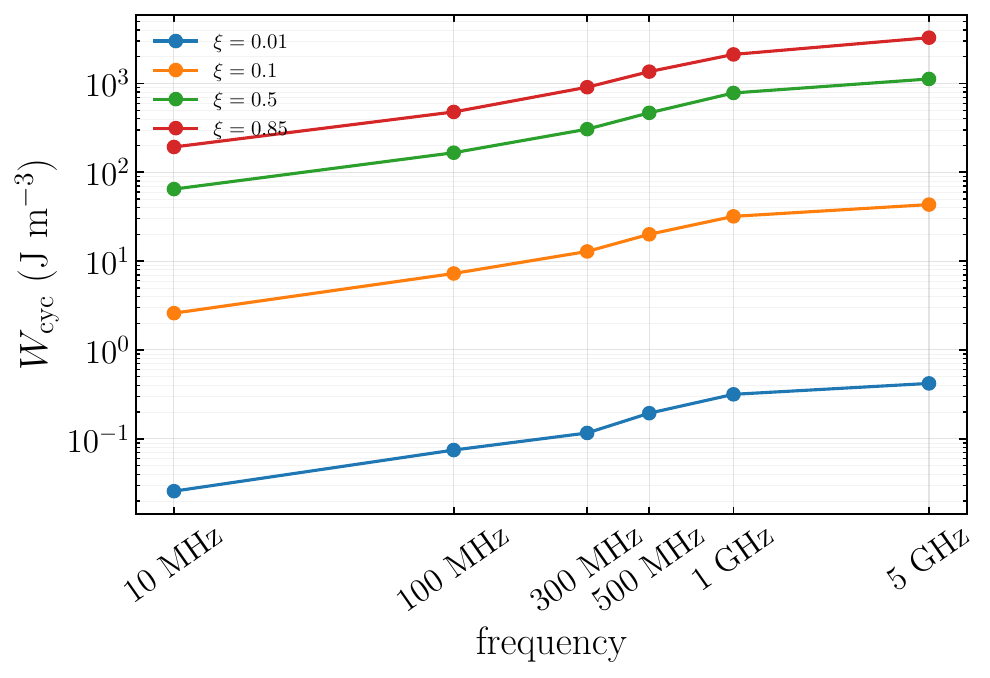}
    \caption{Relaxational magnetic-work density accumulated over one field period, $W_{\mathrm{cyc}}=\mu_0\int_{\mathrm{cyc}}H\,dM$, as a function of frequency for four drive amplitudes. The plotted work is the longitudinal electron--phonon--magnon relaxation contribution of the homogeneous model, not the total experimental core loss.}
    \label{fig:relaxational_work}
    \end{minipage}
\end{figure*}

Fig.\ref{fig:relaxational_work} shows a strong amplitude dependence of the calculated cycle work. For every displayed $\xi$, $W_{\mathrm{cyc}}$ also increases over the sampled range from 10~MHz to 5~GHz. The calculation has, therefore, not yet reached the asymptotic high-frequency regime described in Sec.~\ref{subsec:limits_validation}, where the response amplitude is suppressed strongly enough that the closed-cycle relaxational work must decrease toward zero. The plotted range should be interpreted as the rising and intermediate part of the relaxational response rather than as evidence that the work grows without bound with frequency. The high-frequency limiting behavior derived in Sec.~\ref{subsec:limits_validation} remains a constraint on any extension of the sweep to larger $\omega$.

At sufficiently small field amplitude, Eq.~\eqref{eq:linear_loss} requires the cycle work at fixed frequency to approach quadratic scaling with $H_0$.  Because Eq.~\eqref{eq:xi_definition} gives $H_0\propto\xi$ for the fixed material spectrum used here, the corresponding small-amplitude
expectation is
\[
    W_{\rm cyc}\propto\xi^2
    \qquad (\xi\rightarrow0).
\]
Departure from this quadratic behavior at larger $\xi$ is consistent with the simultaneous loop deformation and growth of higher-harmonic content, for which a single amplitude-independent $\chi''(\omega)$ is no longer sufficient to characterize the response.

\begin{figure*}[!t]
    \centering
    \subfloat[]{\includegraphics[width=0.48\textwidth]{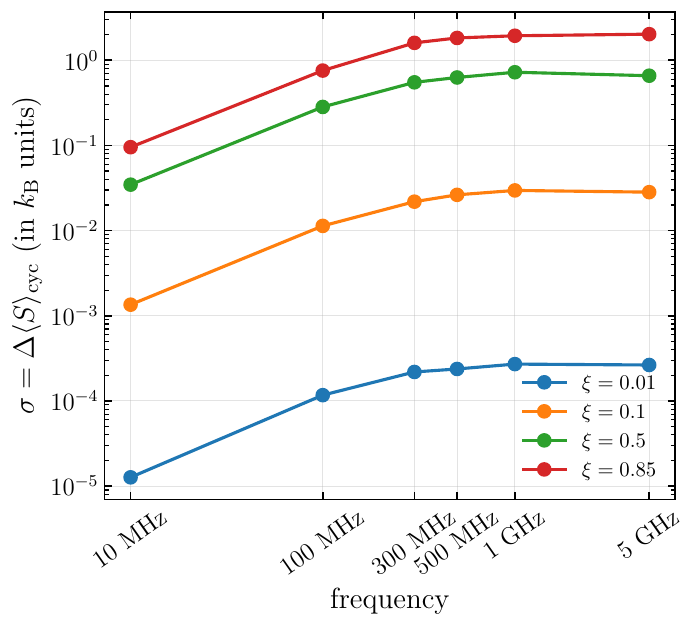}}
    \hfill
    \subfloat[]{\includegraphics[width=0.48\textwidth]{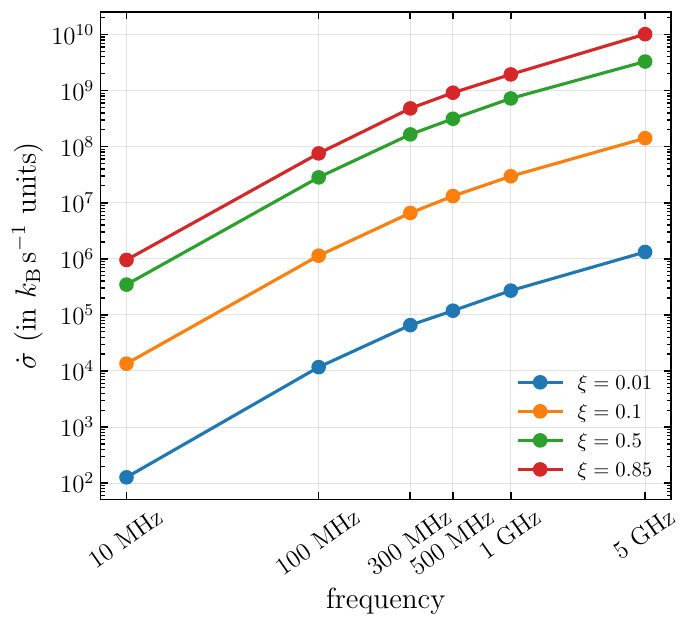}}
    \caption{Entropy production as a function of frequency for four drive amplitudes. Panel (a) shows the cycle-integrated entropy-production measure $\Delta \langle s\rangle_{\mathrm{cyc}}$ in the normalization used by the calculation. Panel (b) shows the corresponding entropy-production-rate output. The two ordinates are reported in the plotting units indicated on the axes and are treated as distinct numerical outputs.}
    \label{fig:entropy_production}
\end{figure*}

The entropy results in Fig.~\ref{fig:entropy_production} provide the complementary thermodynamic measure of irreversibility. The cycle-integrated quantity in panel (a) is nonnegative for every displayed condition and increases strongly with amplitude. With increasing frequency it grows rapidly through the lower and intermediate frequencies and then tends to level off, with a slight decrease at 5~GHz for some of the larger amplitudes. The rate quantity in panel (b), by contrast, continues to increase over the displayed range. These different trends are not contradictory. A per-cycle measure quantifies the irreversible entropy accumulated during one period, while a rate measure additionally weights how quickly such periods occur. The manuscript, therefore, treats the two plotted ordinates according to their definitions and units rather than inferring one from the other.

The nonnegative values are consistent with the SEAQT production law in Eq.~\eqref{eq:entropy_production}, but the logical direction is important as well, namely, that the nonnegativity follows analytically from the quadratic form and is then checked numerically. It is not inferred from the plotted surface. The result also gives a useful distinction between magnetic work and irreversible entropy production. Eq.~\eqref{eq:E0_work_integral} tracks energy transferred into the bare excitation system, whereas Eq.~\eqref{eq:entropy_production} measures the irreversible redistribution of occupation probabilities. There is no general finite-cycle identity $W=T\Delta \langle s\rangle$ for the present isolated node. Such a relation would require additional conditions such as a well-defined externally maintained temperature or a near-isothermal heat-rejection process.

These diagnostics, therefore, serve complementary roles. In the linear periodic limit, $\chi''$ determines the work through Eq.~\eqref{eq:linear_loss}. At finite amplitude, the fundamental quadrature coefficient retains that work connection but becomes amplitude dependent. Entropy production instead measures the irreversibility of the redistribution itself and remains nonnegative whether or not a simple susceptibility description is valid. Using $\chi''$, $W_{\mathrm{cyc}}$, and entropy production together separates phase-lag energetics from the broader thermodynamic statement that the state evolution is irreversible.
\begin{figure*}[!t]
    \centering
    \subfloat[]{\includegraphics[width=0.48\textwidth]{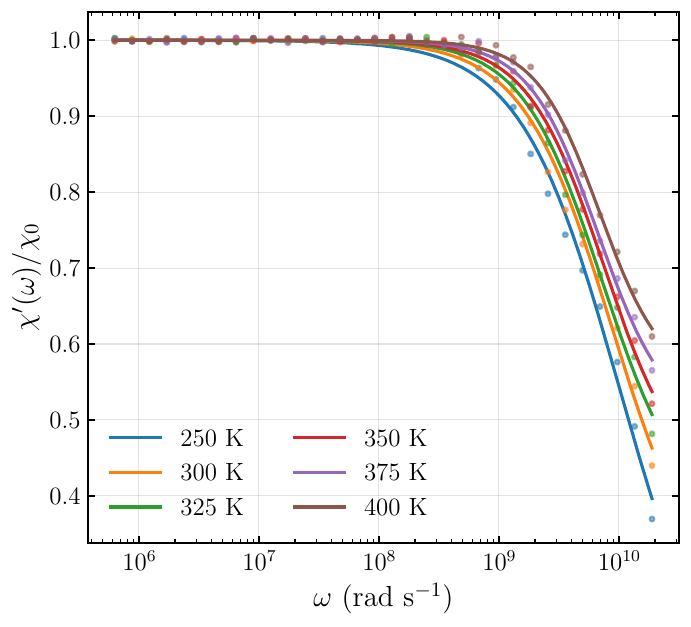}}
    \hfill
    \subfloat[]{\includegraphics[width=0.48\textwidth]{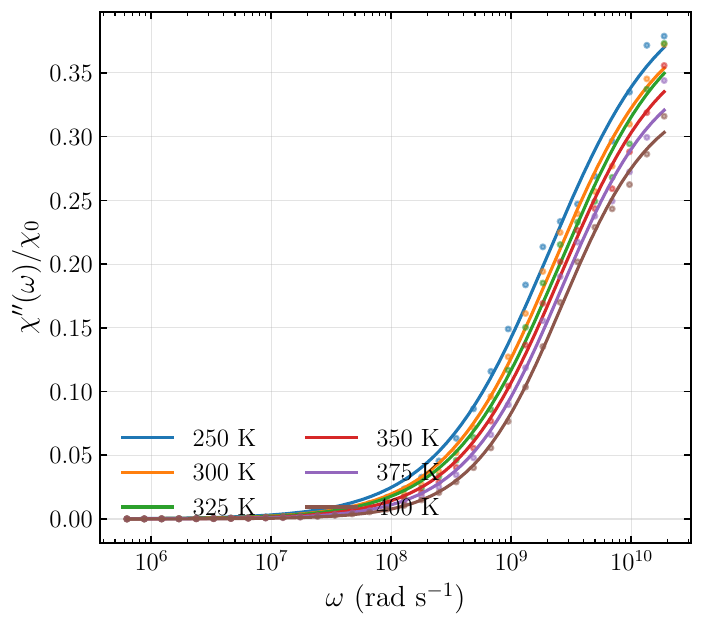}}
    \caption{Temperature dependence of the coupled small-signal susceptibility. Panels (a) and (b) show $\chi'(\omega)$ and $\chi''(\omega)$, respectively, for reference temperatures from 250 to 400~K.}
    \label{fig:temperature_chi}
\end{figure*}
\begin{figure*}[!t]
    \centering
    \subfloat[]{\includegraphics[width=0.32\textwidth]{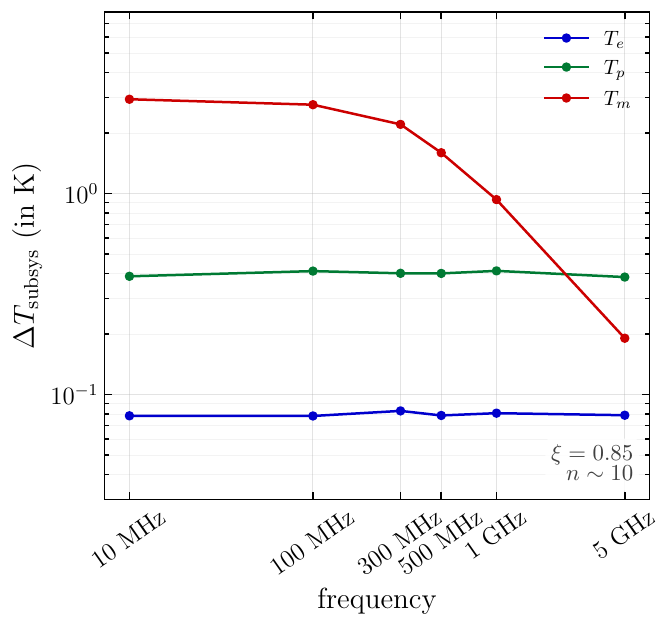}}
    \hfill
    \subfloat[]{\includegraphics[width=0.32\textwidth]{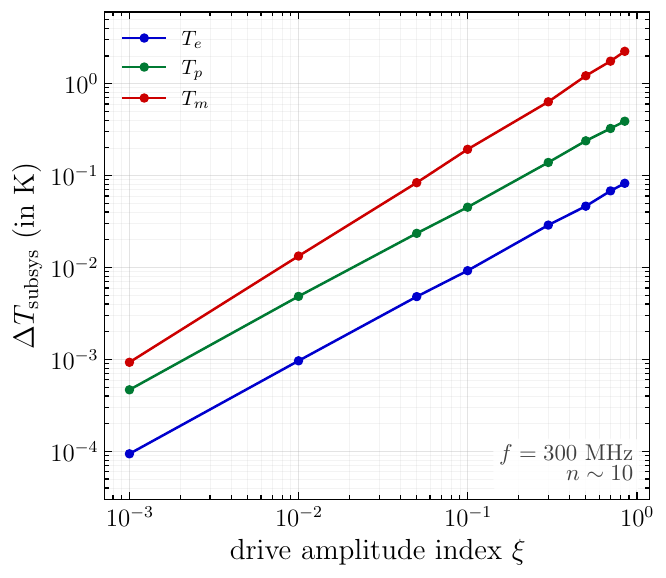}}
    \hfill
    \subfloat[]{\includegraphics[width=0.32\textwidth]{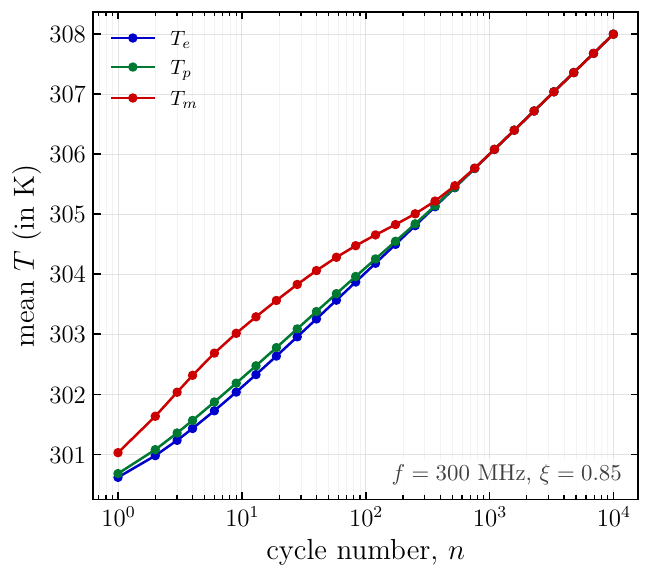}}
    \caption{Non-equilibrium population-temperature response. Panel (a) shows the electron, phonon, and magnon temperature excursions versus frequency at $\xi=0.85$ before substantial secular drift. Panel (b) shows their amplitude dependence at 300~MHz, while panel (c) shows the mean population temperatures over successive cycles at 300~MHz and $\xi=0.85$.}
    \label{fig:subsystem_temperatures}
\end{figure*}

\subsection{Temperature dependence and nonequilibrium population temperatures}
\label{subsec:discussion_temperature}

The complex susceptibility changes systematically with the reference temperature. In Fig.~\ref{fig:temperature_chi}, the low-frequency value of $\chi'$ increases from 250 to 400~K, and the dispersive decrease moves to higher angular frequency as the temperature rises. The loss component $\chi''$ likewise increases in magnitude and its maximum shifts to higher frequency over the displayed interval. These trends arise from the temperature dependence already contained in Eq.~\eqref{eq:coupled_chi} when $\beta_0$ changes directly, while the equilibrium occupation-fluctuation factors $A_i^{k,0}$ redistribute the spectral weights entering $C_e$, $B_{EE}^p$, and the magnon moments. The resulting temperature dependence is, therefore, not imposed through an empirical Curie or Debye law.

This point also defines the limit of the interpretation. The excitation spectra themselves are held fixed in the present calculations. Consequently, Fig.~\ref{fig:temperature_chi} isolates the change produced by thermodynamic occupations and the coupled SEAQT response on those fixed spectra. It does not include explicit temperature-induced renormalization of magnon or phonon dispersions, changes in anisotropy, or critical fluctuations near the Curie temperature. Experimentally, both in-phase and quadrature susceptibilities of magnetite are known to vary with temperature and frequency~\cite{Ozdemir2009}, but the microscopic mechanisms and sample conditions in those measurements need not coincide with the quasiparticle-only channel isolated here. The comparison is, therefore, useful as context for the observables, not as a parameter-free validation of the calculated curves.

Fig.~\ref{fig:subsystem_temperatures}(a) shows how the energy deposited by the field is distributed among the three non-equilibrium population temperatures. Before appreciable secular drift, the magnon population has the largest oscillatory excursion, consistent with its direct coupling to the time-dependent field through $\epsilon_i^m(t)$. The frequency sweep also shows that this magnon-temperature excursion is progressively suppressed toward the highest frequency, paralleling the reduction of the magnetization excursion in Fig.~\ref{fig:mh_frequency}. The phonon excursion remains smaller, and the electron excursion is the smallest over the displayed range. At 300~MHz, all three excursions grow strongly with $\xi$, while retaining the ordering $\Delta T_m>\Delta T_p>\Delta T_e$ [Fig.~\ref{fig:subsystem_temperatures}(b)]. This hierarchy is not imposed by three separate heat baths. It emerges from the combination of direct magnetic driving, spectral response weights, and the shared energy constraint.

The separation of population temperatures has a natural physical interpretation but should not be overmapped onto a particular microscopic scattering time. Time-resolved experiments on Fe$_3$O$_4$ have independently resolved sub-picosecond electron--phonon-associated dynamics and slower spin--lattice or magnetization-recovery components extending to hundreds of picoseconds in nanostructured samples~\cite{Lu2022,Hsia2009}. Those measurements demonstrate that electronic, lattice, and magnetic degrees of freedom in magnetite can exchange energy on distinct time scales, which motivates retaining the three populations explicitly. The effective $\tau_e$, $\tau_p$, and $\tau_m$ used here remain coarse-grained dynamic inputs to the SEAQT metric and are not identified one-to-one with any particular pump--probe decay constant.

Fig.~\ref{fig:subsystem_temperatures}c) makes the absence of heat rejection visible directly. The mean temperatures rise over successive cycles rather than approaching a fixed thermal limit cycle. The three populations initially retain different mean temperatures, but their curves move toward one another as the internally redistributed energy accumulates. This behavior follows the exact energy bookkeeping: when $\Delta \langle e\rangle_{0,\mathrm{cyc}}>0$ and no energy is removed in a heat interaction, the completed thermodynamic state cannot repeat exactly after every cycle even if the fast magnetic trajectory changes only slowly from one period to the next. The calculation, therefore, predicts a separation of time scales between the within-cycle magnetic response and the slower secular evolution of the mean state. A laboratory steady-state temperature would require an external thermal boundary model, specimen geometry, and heat-transfer coefficients and as noted earlier could be modeled within the SEAQT formalism and the hypoequilibrium description with an additional subsystem acting as a thermal reservoir.

\subsection{Integrated physical picture and relation to measured ferrite loss}
\label{subsec:discussion_synthesis}

Taken together, the eight figures presented hereform a consistent causal sequence within the model. The prescribed field first shifts the magnon energy ladder. Finite-rate redistribution then produces a magnon-population lag and, via Eq.~\eqref{eq:M_def}, a lagging longitudinal magnetization. In the linear regime this appears as the coupled complex susceptibility of Fig.~\ref{fig:normalized_chi}. At larger amplitude, the same nonlinear occupation dynamics produces loop-shape distortion and higher harmonics. The common dressed-energy constraint transfers the consequences of the magnetic drive to the phonon and electron blocks, while Eq.~\eqref{eq:entropy_production} requires irreversible redistribution to produce entropy. When the field performs positive net work and no thermal sink is present, that energy remains in the node and appears as the secular temperature drift of Fig.~\ref{fig:subsystem_temperatures}. The magnetic, spectral, energetic, and thermal diagnostics are, therefore, mutually consistent consequences of one state evolution rather than separate phenomenological models fitted to different outputs.

The comparison with conventional ferrite-loss language must, nevertheless, remain precise. In engineering measurements, the area of an observed macroscopic $B$--$H$ or $M$--$H$ loop can contain domain-wall, rotational, eddy-current, dielectric, and other contributions~\cite{Bertotti1988,Lebourgeois1996,Wu2024}. Microwave measurements on magnetite likewise show frequency-dependent magnetic loss and permeability, but those observables can include resonance and specimen-scale electromagnetic effects that are outside the present state space description~\cite{Williams2016}. The work in Fig.~\ref{fig:relaxational_work} is, therefore, not offered as a replacement for measured total core loss. Its role is to isolate a specific intrinsic contribution: irreversible longitudinal redistribution among field-driven magnon states and the coupled electron and phonon populations.

This narrower interpretation is also what makes the framework transferable. Once a consistent electron, phonon, and magnon pseudo-eigenstructure and a dynamic metric are supplied, the same thermodynamic construction can be applied without introducing a new empirical loss law for each material. A subsequent study applies this construction to Fe$_3$O$_4$, MnFe$_2$O$_4$, and (Mn$_{0.5}$Zn$_{0.5}$)Fe$_2$O$_4$ to examine how their spectra and prescribed relaxation scales redistribute the location, breadth, and magnitude of the longitudinal relaxational response~\cite{Dhariwal2026FerriteComparison}. Incorporating the remaining experimental loss channels requires enlarging the physical model---for example by adding spatial electromagnetic transport, domain degrees of freedom, or a heat-rejection boundary---rather than absorbing them into the present quasiparticle work by reinterpretation.

\section{Conclusions}
\label{sec:conclusion}

A field-driven SEAQT description has been formulated for longitudinal relaxation of coupled electron, phonon, and magnon populations in Fe$_3$O$_4$. The magnetic field enters through a shift of the magnon energies, while the occupation basis remains fixed. The irreversible dynamics conserve electron number and instantaneous dressed energy but do not impose magnon-number conservation, allowing the magnon population and longitudinal magnetization to relax. With one effective relaxation parameter per population, the state evolution admits a reduced description in terms of the electron, phonon, and magnon inverse temperatures, the electron-number coordinate, and a field-generated magnon affinity.

The formulation gives two complementary thermodynamic results. First, the entropy production rate is a nonnegative quadratic form in the departures from the instantaneous constrained target. Second, the magnetic work can be followed exactly through the bare-excitation and full field-dependent energy balances. Linearization about an equilibrium reference state produces a coupled complex susceptibility in which all three populations enter through the common energy constraint. A one-pole Debye law is recovered only when the feedback term is negligible. An elliptical $M$--$H$ trajectory is, therefore, a linear-response property, not a general constitutive assumption.

The Fe$_3$O$_4$ calculations illustrate the consequences of that structure. At low frequency the longitudinal response approaches the quasistatic trajectory. Increasing frequency produces substantial phase lag and broad, non-Debye susceptibility dispersion. Increasing field amplitude deforms the $M$--$H$ trajectory away from its first-harmonic ellipse and produces increasing third-harmonic content. Over the sampled 10-MHz--5-GHz interval, the calculated cycle work grows with both frequency and amplitude, while the entropy results distinguish the irreversibility accumulated per cycle from the rate at which it is produced. The susceptibility changes systematically with temperature, and the non-equilibrium temperature calculations show direct magnon heating followed by redistribution to the phonon and electron populations. Without energy rejection via a heat interaction, positive cycle work produces a secular increase of the mean excitation temperature rather than a strict finite-loss thermodynamic limit cycle.

These results should be interpreted within the deliberately restricted model. Domain walls, vector reversal, transverse precession, anisotropy-axis dynamics, spatial eddy currents, demagnetizing-field self-consistency, and external heat transfer are not included. The calculated work is, therefore, the longitudinal quasiparticle-relaxation contribution associated with the coupled electron--phonon--magnon state evolution, not the total core loss of a macroscopic ferrite specimen. Within that scope, the framework provides a direct thermodynamic connection between material excitation spectra, prescribed dynamic time scales, magnetic lag, entropy production, and internal energy redistribution under a time-dependent field.

\appendix

\section{Extension to a network of local material systems}
\label{app:network}

The main derivation and numerical results concern one spatially homogeneous node. This appendix develops a formal extension in which a specimen is partitioned into local material systems connected by inter-node exchange channels. The purpose is twofold: to show how the same SEAQT constraint structure can be applied to spatial energy and particle exchange, and to distinguish the dynamic time scale for transport between nodes from the intra-node relaxation parameters $\tau_e$, $\tau_p$, and $\tau_m$. None of the manuscript figures is generated from this network extension. The construction follows the composite-system and hypoequilibrium logic used previously for interacting SEAQT systems and electron--phonon transport~\cite{LiEnergy2016,Li2018b,BerettaRayVonSpakovsky2026}.

\subsection{Node-resolved state and separation of kinetic processes}
\label{app:node_state}

Let $\ell$ index a local material node of physical volume $\Omega_\ell$. Each node carries its own electronic, phonon, and magnon pseudo-eigenstructures,
\begin{equation}
     \{\varepsilon_i^k,g_i^k,y_i^k,H\}
     \longrightarrow
     \{\varepsilon_i^{k,\ell},g_i^{k,\ell},y_i^{k,\ell},H_\ell\},
     \label{eq:network_mapping}
\end{equation}
with local magnetic shift
\begin{equation}
     h_\ell(t)=\mu_0\mu_m^\ell H_\ell(t),
     \qquad
     \epsilon_i^{m,\ell}(t)=\varepsilon_i^{m,\ell}+h_\ell(t).
     \label{eq:network_field_shift}
\end{equation}
The electronic and phonon energies remain $\epsilon_i^{e,\ell}=\varepsilon_i^{e,\ell}$ and $\epsilon_i^{p,\ell}=\varepsilon_i^{p,\ell}$. Because the DOS quantities in the main text are densities, multiplication by $\Omega_\ell$ converts a nodal density to the corresponding extensive node quantity.

Two types of irreversible process are now distinguished. Intra-node relaxation redistributes energy among the electron, phonon, and magnon populations of one node and is controlled by the local dynamic metric represented in the main text by $\tau_e^\ell$, $\tau_p^\ell$, and $\tau_m^\ell$. Inter-node exchange transfers conserved quantities between neighboring nodes and requires a separate edge metric. Denote a pair of neighboring nodes by $\ell$ and $\ell'$ and, for the simplest reduction below, assign that edge one exchange relaxation parameter $\tau_{\ell\ell'}=\tau_{\ell'\ell}>0$. There is no reason in general for $\tau_{\ell\ell'}$ to equal any of the intra-node relaxation parameters.

The compact energy-only result is obtained under three conditions: the local fields are treated as fixed during the inter-node exchange substep, particles do not cross the edge, and internal equilibration within each node is sufficiently rapid that the node can be represented on the transport time scale by one inverse temperature $\beta^\ell$ with negligible driven magnon affinity. The local electron number remains fixed, so the electronic chemical coordinate adjusts as the node energy changes. These assumptions are relaxed below.

\subsection{Pairwise energy exchange at fixed local electron number}
\label{app:pair_energy}

For a locally equilibrated node, write
\begin{align}
    y_i^{e,\ell}&=\beta^\ell\varepsilon_i^{e,\ell}+\nu^\ell,
     \label{eq:network_local_e}\\
     y_i^{p,\ell}&=\beta^\ell\varepsilon_i^{p,\ell},
     \label{eq:network_local_p}\\
     y_i^{m,\ell}&=\beta^\ell \epsilon_i^{m,\ell}.
     \label{eq:network_local_m}
\end{align}
During exchange along the edge $(\ell,\ell')$, both nodes relax toward a common energy-conjugate multiplier $\beta^{(\ell\ell')}$. Because electron number is conserved separately in each node for this energy-only channel, node $\ell$ also has an edge target $\nu_*^\ell$ chosen to enforce zero electronic number flow. With a common edge time, the instantaneous edge contributions to the occupation rates can be written as
\begin{equation}
     \left.\dot n_i^{k,\ell}\right|_{\ell'}
     =\frac{A_i^{k,\ell}}{\tau_{\ell\ell'}}
     \Delta_i^{k,\ell},
     \label{eq:edge_occupation_rate}
\end{equation}
where
\begin{align}
     \Delta_i^{e,\ell}
     &= (\beta^\ell-\beta^{(\ell\ell')})\varepsilon_i^{e,\ell}
     +(\nu^\ell-\nu_*^\ell),
     \label{eq:edge_delta_e}\\
     \Delta_i^{p,\ell}
     &= (\beta^\ell-\beta^{(\ell\ell')})\varepsilon_i^{p,\ell},
     \label{eq:edge_delta_p}\\
     \Delta_i^{m,\ell}
     &= (\beta^\ell-\beta^{(\ell\ell')})\epsilon_i^{m,\ell}.
     \label{eq:edge_delta_m}
\end{align}
An analogous set holds for node $\ell'$.

Next, the electronic fluctuation moments of node $\ell$ are defined as
\begin{align}
     B_{NN}^{e,\ell}
     &=\sum_i g_i^{e,\ell}A_i^{e,\ell},
     \label{eq:edge_BNN}\\
     B_{EN}^{e,\ell}
     &=\sum_i g_i^{e,\ell}A_i^{e,\ell}\varepsilon_i^{e,\ell},
     \label{eq:edge_BEN}\\
     B_{EE}^{e,\ell}
     &=\sum_i g_i^{e,\ell}A_i^{e,\ell}(\varepsilon_i^{e,\ell})^2.
     \label{eq:edge_BEE}
\end{align}
Separate conservation of electron number in node $\ell$ requires
\begin{equation}
     0=\sum_i g_i^{e,\ell}A_i^{e,\ell}\Delta_i^{e,\ell},
     \label{eq:edge_local_number_constraint}
\end{equation}
which fixes the edge target chemical coordinate as
\begin{equation}
     \nu^\ell-\nu_*^\ell
     =-(\beta^\ell-\beta^{(\ell\ell')})
     \frac{B_{EN}^{e,\ell}}{B_{NN}^{e,\ell}}.
     \label{eq:edge_nu_target}
\end{equation}
Substitution into the electron energy rate removes the number-conserving direction and leaves the fixed-number variance
\begin{equation}
     \mathcal C_e^\ell
     =B_{EE}^{e,\ell}
     -\frac{(B_{EN}^{e,\ell})^2}{B_{NN}^{e,\ell}}
     \ge 0.
     \label{eq:network_Ce}
\end{equation}
For phonons and dressed magnons,
\begin{align}
     \mathcal C_p^\ell
     &=\sum_i g_i^{p,\ell}A_i^{p,\ell}(\varepsilon_i^{p,\ell})^2,
     \label{eq:network_Cp}\\
     \mathcal C_m^\ell
     &=\sum_i g_i^{m,\ell}A_i^{m,\ell}[\epsilon_i^{m,\ell}(t)]^2,
     \label{eq:network_Cm}
\end{align}
and, of course,
\begin{equation}
     \mathcal C^\ell
     =\mathcal C_e^\ell+\mathcal C_p^\ell+\mathcal C_m^\ell.
     \label{eq:network_C}
\end{equation}
These quantities have the same fluctuation structure as the response moments in the main text but do not contain the intra-node factors $1/\tau_k^\ell$. They belong to the edge exchange problem, whose dynamic prefactor is $1/\tau_{\ell\ell'}$.

Taking the energy moment of Eq.~\eqref{eq:edge_occupation_rate} gives the density rate in node $\ell$ due to this one edge,
\begin{equation}
     \left.\dot \langle e\rangle^\ell\right|_{\ell'}
     =\frac{\mathcal C^\ell}{\tau_{\ell\ell'}}
     (\beta^\ell-\beta^{(\ell\ell')}),
     \label{eq:edge_energy_rate}
\end{equation}
where $\langle e\rangle^\ell$ denotes the instantaneous dressed energy density for the fixed local field used in the exchange substep. The corresponding equation for $\ell'$ has the same form. Conservation of the extensive pair energy requires
\begin{equation}
     \Omega_\ell\left.\langle \dot e\rangle^\ell\right|_{\ell'}
     +\Omega_{\ell'}\left.\langle\dot e\rangle^{\ell'}\right|_{\ell}=0.
     \label{eq:pair_energy_conservation}
\end{equation}
Substituting Eq.~\eqref{eq:edge_energy_rate} into Eq.~\eqref{eq:pair_energy_conservation} yields the common pair target
\begin{equation}
     \boxed{
     \beta^{(\ell\ell')}
     =\frac{\Omega_\ell\mathcal C^\ell\beta^\ell
     +\Omega_{\ell'}\mathcal C^{\ell'}\beta^{\ell'}}
     {\Omega_\ell\mathcal C^\ell+\Omega_{\ell'}\mathcal C^{\ell'}}}.
     \label{eq:pair_beta}
\end{equation}
Thus, the interaction target is not an arithmetic average of the two inverse temperatures. Each node is weighted by the amount of thermodynamically active energy fluctuation available to the edge and by the physical volume represented by that node.

The same fluctuation measure gives rise to the local derivative of the energy with respect to the inverse temperature at fixed electron number and fixed local field, i.e.,
\begin{equation}
     \left(\frac{\partial \langle e\rangle^\ell}{\partial\beta^\ell}\right)_{N_e^\ell,H_\ell}
     =-\mathcal C^\ell.
     \label{eq:dE_dbeta_network}
\end{equation}
Eq.~\eqref{eq:edge_energy_rate} is, therefore, equivalent to the reduced inverse-temperature equation
\begin{equation}
     \left.\dot\beta^\ell\right|_{\ell'}
     =-\frac{1}{\tau_{\ell\ell'}}
     (\beta^\ell-\beta^{(\ell\ell')}).
     \label{eq:edge_beta_target}
\end{equation}
Substituting Eq.~\eqref{eq:pair_beta} yields
\begin{align}
     \left.\dot\beta^\ell\right|_{\ell'}
     &=-\frac{1}{\tau_{\ell\ell'}}
     \frac{\Omega_{\ell'}\mathcal C^{\ell'}}
     {\Omega_\ell\mathcal C^\ell+\Omega_{\ell'}\mathcal C^{\ell'}}
     (\beta^\ell-\beta^{\ell'}),
     \label{eq:pair_relax_l}\\
     \left.\dot\beta^{\ell'}\right|_{\ell}
     &=-\frac{1}{\tau_{\ell\ell'}}
     \frac{\Omega_\ell\mathcal C^\ell}
     {\Omega_\ell\mathcal C^\ell+\Omega_{\ell'}\mathcal C^{\ell'}}
     (\beta^{\ell'}-\beta^\ell).
     \label{eq:pair_relax_lp}
\end{align}
The local intra-node relaxation parameters do not appear in these equations. Instead, $\tau_{\ell\ell'}$ is an independent transport relaxation parameter associated with the edge.

It is useful to write the same result as an extensive energy current. Defining positive $\mathcal J^E_{\ell\leftarrow\ell'}$ as energy entering node $\ell$ from node $\ell'$, Eqs.~\eqref{eq:edge_energy_rate} and \eqref{eq:pair_beta} give
\begin{equation}
     \boxed{
     \mathcal J^E_{\ell\leftarrow\ell'}
     =\frac{1}{\tau_{\ell\ell'}}
     \frac{\Omega_\ell\mathcal C^\ell\,
       \Omega_{\ell'}\mathcal C^{\ell'}}
     {\Omega_\ell\mathcal C^\ell+\Omega_{\ell'}\mathcal C^{\ell'}}
     (\beta^\ell-\beta^{\ell'})}.
     \label{eq:edge_energy_current}
\end{equation}
The current is antisymmetric,
\begin{equation}
     \mathcal J^E_{\ell'\leftarrow\ell}
     =-\mathcal J^E_{\ell\leftarrow\ell'},
     \label{eq:edge_current_antisym}
\end{equation}
so every edge conserves total energy independently.

The sign of Eq.~\eqref{eq:edge_energy_current} is thermodynamically correct because a colder node has the larger $\beta$ since $\beta=(k_BT)^{-1}$. If $T_\ell<T_{\ell'}$, then $\beta^\ell>\beta^{\ell'}$ and Eq.~\eqref{eq:edge_energy_current} gives positive energy flow into the colder node. Moreover, at fixed electron number and field, $\partial \langle s\rangle/\partial \langle e\rangle)_{\langle n\rangle_e, H} =k_B\beta$. The entropy production rate associated with this pair is, therefore,
\begin{align}
     \dot \sigma_{\ell\ell'}
     &=k_B(\beta^\ell-\beta^{\ell'})
     \mathcal J^E_{\ell\leftarrow\ell'}
     \nonumber\\
     &=\frac{k_B}{\tau_{\ell\ell'}}
     \frac{\Omega_\ell\mathcal C^\ell\,
       \Omega_{\ell'}\mathcal C^{\ell'}}
     {\Omega_\ell\mathcal C^\ell+\Omega_{\ell'}\mathcal C^{\ell'}}
     (\beta^\ell-\beta^{\ell'})^2
     \ge 0.
     \label{eq:edge_entropy_production}
\end{align}
Thus, the compact pairwise reduction conserves energy and satisfies the second law edge by edge.

The fluctuation measure can also be connected to an ordinary volumetric heat capacity. At fixed $\langle n\rangle_e^\ell$ and $H_\ell$ so that
\begin{equation}
     c_\ell
     \equiv\left(\frac{\partial \langle e\rangle^\ell}{\partial T_\ell}\right)_{\langle n\rangle_e^\ell,H_\ell}
     =\frac{\mathcal C^\ell}{k_BT_\ell^2}
     =k_B(\beta^\ell)^2\mathcal C^\ell.
     \label{eq:network_heat_capacity}
\end{equation}
Eq.~\eqref{eq:pair_beta} is, thus, the inverse-temperature counterpart of a capacity-weighted pair equilibration law.

\subsection{Assembly of a network and the continuum limit}
\label{app:network_assembly}

For a node connected to several neighbors, the pairwise contributions can be superposed over graph edges,
\begin{equation}
     \left.\dot\beta^\ell\right|_{\mathrm{network}}
     =-\sum_{\ell'\sim\ell}
     \frac{1}{\tau_{\ell\ell'}}
     \frac{\Omega_{\ell'}\mathcal C^{\ell'}}
     {\Omega_\ell\mathcal C^\ell+\Omega_{\ell'}\mathcal C^{\ell'}}
     (\beta^\ell-\beta^{\ell'}).
     \label{eq:network_beta}
\end{equation}
The equivalent energy-density form is
\begin{equation}
     \left.\langle \dot e\rangle^\ell\right|_{\mathrm{network}}
     =\frac{1}{\Omega_\ell}
     \sum_{\ell'\sim\ell}
     \mathcal J^E_{\ell\leftarrow\ell'}.
     \label{eq:network_energy_balance}
\end{equation}
Summing Eq.~\eqref{eq:network_energy_balance} over all nodes cancels every edge current with its opposite and, therefore, conserves the total network energy. Likewise, the network entropy production rate is the sum of the nonnegative pair contributions, i.e.,
\begin{equation}
     \dot \sigma_{\mathrm{network}}
     =\sum_{\langle\ell,\ell'\rangle}
     \dot \sigma_{\ell\ell'}\ge0,
     \label{eq:network_entropy_production}
\end{equation}
where each undirected edge is counted once.

For identical nodes with common volume $\Omega$, common response measure $\mathcal C$ and common nearest-neighbor exchange relaxation parameter $\tau_{\mathrm{tr}}$, Eq.~\eqref{eq:network_beta} simplifies to
\begin{equation}
     \dot\beta^\ell
     =\frac{1}{2\tau_{\mathrm{tr}}}
     \sum_{\ell'\sim\ell}(\beta^{\ell'}-\beta^\ell).
 \label{eq:network_equal_nodes}
\end{equation}
On a regular lattice with nearest-neighbor spacing $a$, linearization about a uniform reference state yields
\begin{equation}
     \frac{\partial\beta}{\partial t}
     =\alpha_{\mathrm{tr}}\nabla^2\beta,
     \qquad
     \alpha_{\mathrm{tr}}=\frac{a^2}{2\tau_{\mathrm{tr}}}.
     \label{eq:network_continuum_beta}
\end{equation}
Since a small temperature perturbation satisfies $\delta\beta=-\delta T/(k_BT_0^2)$, the same diffusivity governs the linearized temperature field, namely,
\begin{equation}
     \frac{\partial\delta T}{\partial t}
     =\alpha_{\mathrm{tr}}\nabla^2\delta T.
     \label{eq:network_heat_equation}
\end{equation}
At the uniform reference state used for the linearization, all identical nodes have the same volumetric heat capacity.  This reference-state value is given by
\begin{equation}
     c_0
     \equiv
     \left.c_\ell\right|_{T_\ell=T_0}
     =
     \frac{\mathcal C_0}{k_BT_0^2}
     =
     k_B\beta_0^2\mathcal C_0 .
\end{equation}
The subscript $0$, therefore, denotes evaluation at the homogeneous reference state, rather than a particular network node. Using Eq.~\eqref{eq:network_heat_capacity}, the corresponding thermal conductivity for this linearized regular-lattice discretization is
\begin{equation}
 \kappa=c_0\alpha_{\mathrm{tr}}
 =\frac{c_0a^2}{2\tau_{\mathrm{tr}}}.
 \label{eq:network_kappa}
\end{equation}
Eq.~\eqref{eq:network_kappa} should be read as a mapping between an edge relaxation parameter and a continuum transport coefficient after a spatial discretization has been specified, not as a prediction of $\kappa$ from the local SEAQT spectra alone. Different graph geometries, anisotropic edges, or level-dependent edge metrics modify this mapping.

\subsection{Simultaneous energy and electron-number transport}
\label{app:energy_particle_transport}

If electrons can cross an edge, electron number is no longer conserved separately in each node. The correct pair problem then has two shared generators of the motion---dressed energy and electron number---and two common edge multipliers. This is the network analogue of the two-generator construction used for electron transport in Ref.~\cite{Li2018b} and of the $2\times2$ multiplier problem in Sec.~\ref{subsec:driven_equations}.

For compactness, let $a\in\{\ell,\ell'\}$ label the two nodes and let $\epsilon_i^{k,a}$ denote the instantaneous dressed level energy. Next, the volume-weighted edge moments are defined as
\begin{align}
     \mathcal Q_{EE}
     &=\sum_a\Omega_a\sum_{k,i}g_i^{k,a}A_i^{k,a}(\epsilon_i^{k,a})^2,
     \label{eq:edge_QEE}\\
     \mathcal Q_{EN}
     &=\sum_a\Omega_a\sum_i g_i^{e,a}A_i^{e,a}\varepsilon_i^{e,a},
     \label{eq:edge_QEN}\\
     \mathcal Q_{NN}
     &=\sum_a\Omega_a\sum_i g_i^{e,a}A_i^{e,a},
     \label{eq:edge_QNN}
\end{align}
and
\begin{align}
     \mathcal R_E
    &=\sum_a\Omega_a\sum_{k,i}g_i^{k,a}A_i^{k,a}\epsilon_i^{k,a}y_i^{k,a},
     \label{eq:edge_RE}\\
     \mathcal R_N
     &=\sum_a\Omega_a\sum_i g_i^{e,a}A_i^{e,a}y_i^{e,a}.
     \label{eq:edge_RN}
\end{align}
For one common edge relaxation parameter, the kinetic prefactor cancels from the constraint equations and the shared edge targets satisfy
\begin{equation}
     \boxed{
     \begin{pmatrix}
      \mathcal Q_{EE} & \mathcal Q_{EN}\\
      \mathcal Q_{EN} & \mathcal Q_{NN}
     \end{pmatrix}
     \begin{pmatrix}
      \beta^{(\ell\ell')}\\
      \nu^{(\ell\ell')}
     \end{pmatrix}
     =
     \begin{pmatrix}
      \mathcal R_E\\
      \mathcal R_N
     \end{pmatrix}}.
     \label{eq:edge_two_generator_solve}
\end{equation}
This has the same Gram-matrix structure as Eq.~\eqref{eq:two_by_two_solve}. Its determinant is nonnegative by the same weighted-variance argument used in Eq.~\eqref{eq:determinant_positive}. It is strictly positive whenever the active energy direction is independent of the electron-number direction.

Now, defining
\begin{equation}
     \Delta_i^{k,a}
     =y_i^{k,a}-\beta^{(\ell\ell')}\epsilon_i^{k,a}
     -\nu^{(\ell\ell')}\delta_{ke}.
     \label{eq:edge_general_delta}
\end{equation}
and obtaining the shared targets from Eq.~\eqref{eq:edge_two_generator_solve}, the extensive energy and electron-number currents into node $\ell$ are found from
\begin{align}
     \mathcal J^E_{\ell\leftarrow\ell'}
     &=\frac{\Omega_\ell}{\tau_{\ell\ell'}}
     \sum_{k,i}g_i^{k,\ell}A_i^{k,\ell}\epsilon_i^{k,\ell}
     \Delta_i^{k,\ell},
     \label{eq:edge_general_energy_current}\\
     \mathcal J^N_{\ell\leftarrow\ell'}
     &=\frac{\Omega_\ell}{\tau_{\ell\ell'}}
     \sum_i g_i^{e,\ell}A_i^{e,\ell}
     \Delta_i^{e,\ell}.
     \label{eq:edge_general_particle_current}
\end{align}
The two constraint rows in Eq.~\eqref{eq:edge_two_generator_solve} guarantee that the corresponding currents into node $\ell'$ are the negatives of Eqs.~\eqref{eq:edge_general_energy_current} and \eqref{eq:edge_general_particle_current}. The edge entropy production rate is again a nonnegative quadratic form given by
\begin{equation}
     \frac{\dot \sigma_{\ell\ell'}}{k_B}
     =\frac{1}{\tau_{\ell\ell'}}
     \sum_a\Omega_a\sum_{k,i}
     g_i^{k,a}A_i^{k,a}(\Delta_i^{k,a})^2
     \ge0.
     \label{eq:edge_general_entropy}
\end{equation}
Thus, energy and particle transport can be added without introducing phenomenological Onsager coefficients at the level of the state equation. The transport metric is carried instead by the edge kinetic operator. A level- or species-dependent edge metric is incorporated by replacing the common $1/\tau_{\ell\ell'}$ with the corresponding positive weights throughout the Gram matrix and currents.

\subsection{Time-dependent fields, nonequilibrated nodes, and scope of the extension}
\label{app:network_scope}

The reductions above hold at fixed instantaneous local fields during the transport substep. If $H_\ell(t)$ varies concurrently, external magnetic work and inter-node transport must both appear in the nodal balance. For the full local field-dependent material energy density $\langle e\rangle_{fd_{\ell}}$, the natural bookkeeping is
\begin{equation}
     \langle \dot e\rangle_{fd_{\ell}}
     =-\mu_0M_\ell\dot H_\ell
     +\frac{1}{\Omega_\ell}
     \sum_{\ell'\sim\ell}\mathcal J^E_{\ell\leftarrow\ell'},
     \label{eq:network_driven_energy_balance}
\end{equation}
with additional particle-current terms already included in $\mathcal J^E$ when the two-generator edge problem is used. Eq.~\eqref{eq:network_driven_energy_balance} separates energy supplied by the prescribed field from energy transported between material nodes.

If the three populations within a node are not internally equilibrated on the transport time scale, one scalar $\beta^\ell$ is insufficient. The node must then retain its local variables $\{\beta_e^\ell,\nu_e^\ell,\beta_p^\ell,\beta_m^\ell,\eta_m^\ell\}$ or, more generally, the full level variables $y_i^{k,\ell}$. Inter-node exchange can still be constructed by applying the same Gram-matrix procedure to the generators carried by a particular edge. The compact Eqs.~\eqref{eq:pair_beta}--\eqref{eq:network_kappa} are, therefore, a reduced transport limit, not a restriction on the more general network formulation.

The homogeneous limit follows automatically. If all nodes have identical spectra, fields, intensive coordinates, and conserved quantities, every edge departure vanishes and no inter-node current is generated. Conversely, spatial gradients in temperature, electron chemical coordinate, composition, or local field create edge departures and, thus, transport. A spatial electronic-conduction problem requires the energy--number construction of Sec.~\ref{app:energy_particle_transport}. A spatial magnon or spin-transport problem would require an edge variable representing the transported magnetic quantity together with local source and sink terms for processes that do not conserve magnon number.

Finally, a network of thermodynamic nodes is not by itself a complete eddy-current or magnetostatic model. The local field $H_\ell(t)$ used by each node must be supplied consistently. If the applied field is modified by demagnetizing fields, induced currents, or electromagnetic propagation, the network SEAQT equations must be coupled to the appropriate Maxwell problem and specimen geometry. The extension derived here, therefore, supplies the thermodynamic state and exchange algebra needed for spatial coupling while leaving the electromagnetic closure as a separate physical model.

\section{Derivation of linearized dressed-energy constraint and the susceptibility}
\label{app:susceptibility_derivation}

This appendix provides the intermediate algebra leading to the linearized common energy-multiplier response and the coupled susceptibility given in Sec.~\ref{subsec:linear_response}. The derivation is carried out about the equilibrium reference state
defined in Eq.~\eqref{eq:reference_state}.

For any first-order quantity $\delta x(t)$, the complex harmonic amplitude is denoted here by a breve,
\begin{equation}
     \delta x(t)
     =
     \operatorname{Re}
     \left[
     \delta\breve{x}\,e^{i\omega t}
     \right].
     \label{eq:app_breve_convention}
\end{equation}
Thus,
\begin{equation}
     \frac{d}{dt}\delta x(t)
     \longleftrightarrow
     i\omega\,\delta\breve{x}.
     \label{eq:app_harmonic_derivative}
\end{equation}
The breve, therefore, denotes a complex first-order harmonic amplitude and is distinct from the hat notation used elsewhere for operators.
In particular,
\begin{equation}
     \delta\breve{h}
     =
     \mu_0\mu_m\,\delta\breve{H}.
     \label{eq:app_dh_dH}
\end{equation}
Furthermore, at the reference state,
\begin{equation}
     \bar \epsilon_i^m=\varepsilon_i^m+\bar h,
     \qquad
     \Delta_i^{k,0}=0,
     \label{eq:app_reference_departure}
\end{equation}
and the equilibrium occupation-fluctuation factors are denoted by $A_i^{k,0}$.

\subsection{Linearized level departures}

Linearization of Eqs.~\eqref{eq:y_e_eom}--\eqref{eq:y_m_eom}
about the equilibrium reference state yields
\begin{align}
     D_e\,\delta\breve{y}_i^e
     &=
     \varepsilon_i^e\,\delta\breve{\beta}
     +\delta\breve{\nu},
     \label{eq:app_lin_ye}\\
     D_p\,\delta\breve{y}_i^p
     &=
     \varepsilon_i^p\,\delta\breve{\beta},
     \label{eq:app_lin_yp}\\
     D_m\,\delta\breve{y}_i^m
     &=
     \bar \epsilon_i^m\,\delta\breve{\beta}
     +\beta_0\,\delta\breve{h}.
     \label{eq:app_lin_ym}
\end{align}
where 
\begin{equation}
     D_k(\omega)\equiv 1+i\omega\tau_k,
     \qquad k\in\{e,p,m\}.
     \label{eq:app_Dk}
\end{equation}
For the magnon block, the term proportional to
$\delta\breve{h}$ follows from the first-order variation
\begin{equation}
     \delta(\beta \epsilon_i^m)
     =
     \bar \epsilon_i^m\,\delta \breve \beta
     +
     \beta_0\,\delta \breve h,
     \label{eq:app_beta_w_variation}
\end{equation}
where the product $\delta \breve \beta\,\delta \breve h$ is second order and is, therefore, omitted.

The departures from the instantaneous constrained target are defined in Eq.~\eqref{eq:Delta_def}.  Their first-order complex amplitudes are 
\begin{align}
     \delta\breve{\Delta}_i^e
     &=
     \delta\breve{y}_i^e
     -\varepsilon_i^e\,\delta\breve{\beta}
     -\delta\breve{\nu},
     \label{eq:app_dDelta_e_start}\\
     \delta\breve{\Delta}_i^p
     &=
     \delta\breve{y}_i^p
     -\varepsilon_i^p\,\delta\breve{\beta},
     \label{eq:app_dDelta_p_start}\\
     \delta\breve{\Delta}_i^m
     &=
     \delta\breve{y}_i^m
     -\bar \epsilon_i^m\,\delta\breve{\beta}
     -\beta_0\,\delta\breve{h}.
     \label{eq:app_dDelta_m_start}
\end{align}
Substitution of Eqs.~\eqref{eq:app_lin_ye}--\eqref{eq:app_lin_ym} gives
\begin{align}
     \delta\breve{\Delta}_i^e
     &=
     -\frac{i\omega\tau_e}{D_e}
     \left(
     \varepsilon_i^e\,\delta\breve{\beta}
     +\delta\breve{\nu}
     \right),
     \label{eq:app_dDelta_e}\\
     \delta\breve{\Delta}_i^p
     &=
     -\frac{i\omega\tau_p}{D_p}
     \varepsilon_i^p\,\delta\breve{\beta},
     \label{eq:app_dDelta_p}\\
     \delta\breve{\Delta}_i^m
     &=
     -\frac{i\omega\tau_m}{D_m}
     \left(
     \bar \epsilon_i^m\,\delta\breve{\beta}
     +\beta_0\,\delta\breve{h}
     \right).
     \label{eq:app_dDelta_m}
\end{align}

\subsection{Electron-number constraint}

The exact electron-number constraint is Eq.~\eqref{eq:constraint_N},
\begin{equation}
     \sum_i r_i^e\Delta_i^e=0,
\end{equation}
with
\begin{equation}
     r_i^e=\frac{g_i^eA_i^e}{\tau_e}.
\end{equation}
Its first-order variation about equilibrium is
\begin{equation}
     \delta
     \left(
     \sum_i r_i^e\Delta_i^e
     \right)
     =
     \sum_i
     \left[
     r_i^{e,0}\delta\Delta_i^e
     +
     \delta r_i^e\,\Delta_i^{e,0}
     \right]
     =0.
     \label{eq:app_N_constraint_variation}
\end{equation}
Because $\Delta_i^{e,0}=0$, the terms containing $\delta r_i^e$ vanish.  Hence
\begin{equation}
     \sum_i
     \frac{g_i^eA_i^{e,0}}{\tau_e}
     \delta\breve{\Delta}_i^e
     =0.
     \label{eq:app_N_constraint_linear}
\end{equation}
Using Eq.~\eqref{eq:app_dDelta_e} then yields
\begin{equation}
     -\frac{i\omega}{D_e}
     \left(
     B_{EN}^e\,\delta\breve{\beta}
     +
     B_{NN}^e\,\delta\breve{\nu}
     \right)
     =0.
     \label{eq:app_N_constraint_moments}
\end{equation}
For a finite-frequency harmonic perturbation,
\begin{equation}
     B_{EN}^e\,\delta\breve{\beta}
     +
     B_{NN}^e\,\delta\breve{\nu}
     =0,
\end{equation}
and, therefore,
\begin{equation}
     \boxed{
     \delta\breve{\nu}
     =
     -\frac{B_{EN}^e}{B_{NN}^e}
     \delta\breve{\beta}}.
     \label{eq:app_delta_nu}
\end{equation}
This is the electron-number relation used in the main-text linear-response derivation.

\subsection{Linearized dressed-energy constraint}

The exact dissipative dressed-energy constraint is Eq.~\eqref{eq:constraint_E},
\begin{equation}
     \sum_{k,i}r_i^k \epsilon_i^k\Delta_i^k=0.
\end{equation}
Its first-order variation is
\begin{align}
    0
    &=
    \delta
    \left(
    \sum_{k,i}r_i^k \epsilon_i^k\Delta_i^k
    \right)
    \nonumber\\
    &=
    \sum_{k,i}
    \left[
    r_i^{k,0}\bar \epsilon_i^k\,\delta\Delta_i^k
    +
    \delta(r_i^k \epsilon_i^k)\Delta_i^{k,0}
    \right],
\end{align}
where
\begin{equation}
     \bar \epsilon_i^e=\varepsilon_i^e,\qquad
     \bar \epsilon_i^p=\varepsilon_i^p,\qquad
     \bar \epsilon_i^m=\varepsilon_i^m+\bar h.
\end{equation}
Again, $\Delta_i^{k,0}=0$, so all first-order variations of the response weights and dressed energies multiplying the reference departure vanish from the constraint.  As a consequence,
\begin{equation}
     \sum_{k,i}
     \frac{g_i^kA_i^{k,0}}{\tau_k}
     \bar \epsilon_i^k
     \delta\breve{\Delta}_i^k
     =0.
     \label{eq:app_E_constraint_linear}
\end{equation}

Substituting Eqs.~\eqref{eq:app_dDelta_e}--\eqref{eq:app_dDelta_m} gives
\begin{align}
    0=-i\omega\Bigg[
    &
    \frac{
     B_{EE}^e\,\delta\breve{\beta}
     +B_{EN}^e\,\delta\breve{\nu}}
     {D_e}
    \nonumber\\
    &+
    \frac{
     B_{EE}^p\,\delta\breve{\beta}}
     {D_p}
    \nonumber\\
    &+
    \frac{
     B_{EE}^m\,\delta\breve{\beta}
     +\beta_0B_{EN}^m\,\delta\breve{h}}
     {D_m}
    \Bigg].
     \label{eq:app_E_constraint_expanded}
\end{align}
For $\omega\neq0$, the common factor $-i\omega$ can be removed.  Inserting Eq.~\eqref{eq:app_delta_nu} into the electronic part then results in
\begin{align}
     B_{EE}^e\,\delta\breve{\beta}
     +B_{EN}^e\,\delta\breve{\nu}
     &=
     \left[
     B_{EE}^e
     -\frac{(B_{EN}^e)^2}{B_{NN}^e}
     \right]
     \delta\breve{\beta}
     \nonumber\\
     &=
     C_e\,\delta\breve{\beta}.
     \label{eq:app_electron_reduction}
\end{align}
The dressed-energy constraint, therefore, becomes
\begin{align}
    0=&
    \left[
     \frac{C_e}{D_e}
     +\frac{B_{EE}^p}{D_p}
     +\frac{B_{EE}^m}{D_m}
    \right]
    \delta\breve{\beta}
    \nonumber\\
    &+
    \frac{\beta_0B_{EN}^m}{D_m}
    \delta\breve{h}.
     \label{eq:app_E_constraint_reduced}
\end{align}

Substituting Eq. (\ref{eq:Gomega}), this last expression is written as
\begin{equation}
     G(\omega)\,\delta\breve{\beta}
     +
     \frac{\beta_0B_{EN}^m}
     {1+i\omega\tau_m}
     \delta\breve{h}
     =0.
     \label{eq:app_delta_beta_intermediate}
\end{equation}
Solving for the common energy-multiplier amplitude then yields
\begin{equation}
     \boxed{
     \delta\breve{\beta}
     =
     -\frac{
     \beta_0\delta\breve{h}\,B_{EN}^m}
     {(1+i\omega\tau_m)G(\omega)}}.
     \label{eq:app_delta_beta}
\end{equation}
This reproduces Eq.~\eqref{eq:delta_beta} of the main text. The result is obtained for a finite-frequency harmonic perturbation. The quasistatic result follows from the continuous $\omega\rightarrow0$ in the limit.

\subsection{Derivation of the coupled susceptibility}

The magnetization is related to the total magnon population by Eq.~\eqref{eq:M_def}. Its first-order variation is, therefore,
\begin{equation}
     \delta M=-\mu_m\,\delta \langle n\rangle_m.
     \label{eq:app_deltaM_start}
\end{equation}
Since
\begin{equation}
     \langle n\rangle_m=\sum_i g_i^m \langle n\rangle_i^m
\end{equation}
and
\begin{equation}
    \delta \langle n\rangle_i^m
     =
     -A_i^{m,0}\delta y_i^m,
\end{equation}
Eq.~\eqref{eq:app_deltaM_start} becomes
\begin{equation}
     \delta\breve{M}
     =
     \mu_m
     \sum_i
     g_i^mA_i^{m,0}
     \delta\breve{y}_i^m.
     \label{eq:app_deltaM_from_y}
\end{equation}
Using the linearized magnon equation
Eq.~\eqref{eq:app_lin_ym},
\begin{equation}
     \delta\breve{y}_i^m
     =
     \frac{
     \bar \epsilon_i^m\,\delta\breve{\beta}
     +\beta_0\,\delta\breve{h}}
     {1+i\omega\tau_m},
\end{equation}
Eq. (\ref{eq:app_deltaM_from_y}) is written as
\begin{align}
     \delta\breve{M}
     &=
 \frac{\mu_m}{1+i\omega\tau_m}
     \sum_i
     g_i^mA_i^{m,0}
     \left(
     \bar \epsilon_i^m\,\delta\breve{\beta}
     +\beta_0\,\delta\breve{h}
     \right)
     \nonumber\\
     &=
     \frac{\mu_m}{1+i\omega\tau_m}
     \left[
     B_{EN}^m\,\delta\breve{\beta}
     +
     \beta_0B_{NN}^m\,\delta\breve{h}
     \right].
     \label{eq:app_deltaM_before_beta}
\end{align}
Substitution of Eq.~\eqref{eq:app_delta_beta} results in
\begin{align}
     \delta\breve{M}
     &=
     \frac{\mu_m}{1+i\omega\tau_m}
     \left[
     -\frac{
     \beta_0\delta\breve{h}(B_{EN}^m)^2}
     {(1+i\omega\tau_m)G(\omega)}
     +
     \beta_0B_{NN}^m\delta\breve{h}
     \right]
     \nonumber\\
     &=
     \mu_m\beta_0\delta\breve{h}
     \left[
     \frac{B_{NN}^m}{1+i\omega\tau_m}
     -
     \frac{(B_{EN}^m)^2}
     {(1+i\omega\tau_m)^2G(\omega)}
     \right].
     \label{eq:app_deltaM_final}
\end{align}

Finally, using Eq.~\eqref{eq:app_dh_dH} and  
\begin{equation}
     \delta\breve{M}
     =
     \chi(\omega)\,\delta\breve{H}.
     \label{eq:app_chi_definition}
\end{equation}
the complex susceptibility is found to be
\begin{equation}
     \boxed{
     \chi(\omega)
     =
     \mu_0\mu_m^2\beta_0
     \left[
     \frac{B_{NN}^m}{1+i\omega\tau_m}
     -
     \frac{(B_{EN}^m)^2}
     {(1+i\omega\tau_m)^2G(\omega)}
     \right]}.
     \label{eq:app_coupled_chi}
\end{equation}
This reproduces Eq.~\eqref{eq:coupled_chi} of the main text.

The two terms in the square brackets of Eq.~\eqref{eq:app_coupled_chi} have distinct origins.  The first is the direct longitudinal response of the magnon population to the field-induced shift of the magnon energies. The second arises because the field-driven change in magnon energy cannot occur independently of the global dressed-energy constraint. The resulting variation of the common multiplier $\beta$ feeds back onto the magnon occupations.  Via $G(\omega)$, this feedback contains the thermodynamically active electron, phonon, and magnon energy moments together with their respective relaxation parameters. Thus, the second term is not an additional phenomenological correction to a Debye susceptibility. Instead, it follows directly from the coupled SEAQT dressed-energy constraint.

\bibliography{references}
\end{document}